\documentclass[manuscript, nonacm]{acmart}

\newcommand{\remove}[1]{}
\newcommand{\add}[1]{#1}

\AtBeginDocument{%
  }

\copyrightyear{2026}
\acmYear{2026}
\setcopyright{acmlicensed}
\acmConference[CSCW '26]{ACM SIGCHI Conference on Computer-Supported Cooperative Work \& Social Computing}{October 10-14, 2026}{Salt Lake City, USA}
\acmBooktitle{Proceedings of the 2026 Computer-Supported Cooperative Work and Social
Computing (CSCW '26), October 10-14, 2026, Salt Lake City, USA}
\acmDOI{10.1145/XXXXXXX.XXXXXXX}
\acmISBN{978-1-4503-XXXX-X/2026/10}

\author{Long Ling}
\orcid{0009-0001-2635-788X}
\affiliation{
  \institution{Tongji University}
  \department{College of Design and Innovation}
  \city{Shanghai}
  \country{China}
}
\email{lucyling0224@gmail.com}

\author{Xiyu Zheng}
\orcid{0009-0001-2370-1200}
\affiliation{
  \institution{Tongji University}
  \department{College of Design and Innovation}
  \city{Shanghai}
  \country{China}
}
\email{zhengxiyu1019@gmail.com}

\author{Gengchen Cao}
\orcid{0009-0004-6789-2098}
\affiliation{
  \institution{Tsinghua-Anta Joint Research Center}
  \city{Beijing}
  \country{China}
}
\email{gengchencao@gmail.com}

\author{RAY LC}
\authornote{Correspondences can be addressed to ray.lc@cityu.edu.hk.}
\email{ray.lc@cityu.edu.hk}
\orcid{0000-0001-7310-8790}
\affiliation{
\institution{City University of Hong Kong\\Studio for Narrative Spaces}
\city{Hong Kong, SAR}
\country{China}}

\begin{document}

\title["Re-Tell the Fortune so I Can Believe It"]{"Re-Tell the Fortune so I Can Believe It": How Chinese User Communities Engage with and Interpret GenAI-based Fortune-Telling}

\renewcommand{\shortauthors}{Long, et al.}

\begin{abstract}
People traditionally divine the future by interpreting natural phenomena as oracular signals, especially in societies adhering to traditional beliefs like China. With the advent of Generative AI (GenAI), people gain access to new ways of probing digital oracles for predicting the future. To understand how people use and interpret GenAI for divination in China, we interviewed 22 participants who habitually use GenAI platforms for fortune-telling, complemented by a three-week digital ethnography with 1,842 community posts. Qualitative analysis showed that people who seek psychological comfort are particularly receptive to GenAI-based decision-making. Users valued GenAI's accessibility, convenience, and efficiency while perceiving its lack of spiritual mystique. We observed community dynamics forming around GenAI tools, where users reinforce interpretations by sharing and discussing with each other, repeating queries until responses align with expectations. Our work uncovers how AI technologies change the way people and communities engage in traditional cultural practices while yearning for the same goals.
\end{abstract}



\received{13 May 2025}
\received[accepted]{18 March 2026}

\maketitle

\section{Introduction}\label{sec:Introduction}
Fortune-telling, or divination, has deep historical roots in Chinese communities, functioning not only as a means of predicting outcomes but also as a cultural practice for navigating uncertainty and maintaining social bonds. Traditional local methods are primarily centered on the interpretation of natural and metaphysical symbols, from oracle bone divination in China's Shang Dynasty to other practices such as \textit{BaZi} (the Four Pillars of Destiny), \textit{Ziwei Doushu} (Purple Star Astrology), \textit{Feng Shui} (Chinese Geomancy), \textit{Physiognomy} (Face Reading), and \textit{Oneiromancy} (Dream Interpretation). Regardless of the specific methodology used, consulting fortune-tellers remains a prevalent practice in Chinese communities. It reflects a persistent quest for spiritual guidance, emotional stability, and decisional clarity when navigating uncertainties in relationships, career changes, and business ventures~\cite{Matthews2020TheYE}. Meanwhile, imported divinatory systems like tarot and astrology are gaining increasing popularity among younger generations~\cite{Tsang2004TowardAS, Matthews2020TheYE}. The ongoing interplay between these distinct traditions has fostered the emergence of a diverse Chinese spiritual divination landscape. 

More recently, Chinese communities have absorbed new forms of divination through the adoption of generative AI (GenAI)~\cite{reed2021ai}. Thanks to its conversational and generative capacities~\cite{yang_ai_2022,han_when_2024,he_i_2025}, GenAI can imitate the listening and dialogic styles of human fortune-tellers, making it increasingly common for people to turn to AI for guidance~\cite{lustig2022explainability}. Importantly, these systems are not simply transplanted technologies but are culturally localized. Large language models (LLMs) such as ChatGPT and DeepSeek are adapted to perform practices with strong Chinese cultural significance, including \textit{BaZi} readings, palmistry, and physiognomy. Beyond mere simulation, the use of AI divination introduces new interactive features and social affordances, giving rise to emergent community ecologies. For instance, Cece, a widely used Chinese fortune-telling platform, allows users to exchange divinatory outcomes, debate interpretations, and collectively reflect on spiritual meanings within online spaces. At their core, both fortune-telling and GenAI aim to impose order on uncertainty by uncovering hidden patterns and generating predictions. While traditional divination is rooted in natural symbolic interpretation and metaphysical worldviews, GenAI operates through data-driven inference and statistical modeling. 

Although their forms may appear similar, the underlying trust logic has shifted. Unlike traditional divination, where authority resides in the practitioner and is reinforced by communal norms, GenAI systems shift trust onto models that lack moral judgment and community accountability, creating both novel affordances and potential risks. The ease and accessibility of GenAI readings can encourage overreliance, altering spiritual practices, social interactions, and everyday decision-making. These dynamics are particularly significant in China, where divination remains a culturally rooted, communal practice, and where GenAI adoption reshapes emotional, social, and spiritual engagement. Understanding this shift is also critical for research on algorithmically mediated communities, culturally situated GenAI~\cite{huang_not_2026}, and the interplay of trust, emotion, and social validation.

Growing attention has been paid to the emotional and spiritual dimensions of human-AI interaction. Prior work has explored how religious leaders adopt digital tools in spiritual practices~\cite{10.1145/1180875.1180908, simmerlein2024sacred, campbell2017surveying}, how visualization tools can strengthen spiritual support networks in health contexts~\cite{10.1145/3462204.3481774}, and how users attribute spiritual authority to AI agents, such as in the cases of the robotic Buddhist priest Mindar and the Spirituality Chatbot~\cite{LoewenColn2022FabulationMA}, demonstrating how AI transforms emotionally-driven and spiritual interactions. Moreover, Graves~\cite{Graves2021EmergentMF} has also called for an emergentist view of AI spirituality, highlighting how human-AI spiritual interaction challenges traditional boundaries between computation and transcendence. While these studies demonstrate that AI can transform spiritual interactions, they primarily focus on individual-agent dyads or the digitization of established religious roles. However, the emergence of GenAI-based fortune-telling introduces a more complex, socially mediated interaction logic. Crucially, little is known about how users engage with GenAI-based fortune-telling and how these spiritual experiences shape their trust and usage patterns, especially in culturally specific contexts like China, where divination remains both personal and collective. Understanding these user interactions is critical, as it reveals how traditional belief systems and modern technology merged, offering broader insights into how trust is renegotiated when AI enters highly sensitive, culturally-rooted social spheres.

Therefore, we focus on how individuals and their broader social circles interact with, trust, and attribute meaning to these GenAI readings through the following research questions:
 
\textbf{RQ1}: \textit{What are the key motivations and interaction practices when Chinese users engage with GenAI-based fortune-telling?}

\textbf{RQ2}: \textit{How do users build trust in GenAI prediction apps and share the belief in these predictions with their community?}

\textbf{RQ3}: \textit{How do users' social relationships with others exposed to GenAI fortune-telling tools affect the way they perceive and use these tools?}

To answer these questions, we first conducted semi-structured interviews with 22 participants representing a wide spectrum of experience with GenAI-based fortune-telling. Based on the communities most frequently mentioned in the interviews, we also conducted a three-week digital ethnography with 1,842 posts to gain more contextualized and authentic insights into community interactions. This involved observing online GenAI fortune-telling communities on WeChat, RedNote, and Cece, focusing on how users engaged with divinatory content in social settings. These observations offered insight into emerging norms, vocabularies, and cultural meanings surrounding AI-mediated divination practices in China. Through qualitative analysis, three researchers independently coded the interview and online ethnography data and iteratively developed themes through collaborative discussion.

Our findings reveal that users’ trust in and acceptance of GenAI-based fortune-telling tools vary significantly depending on their motivations. Those who seek entertainment or psychological comfort tend to embrace these tools more readily, whereas individuals making major life decisions remain cautious, often treating AI predictions as supplementary input rather than authoritative guidance. Users valued GenAI’s accessibility, convenience, and efficiency while perceiving its lack of spiritual mystique. Surprisingly, Chinese users often view the absence of human judgment as a privacy advantage rather than a limitation. However, this absence of ethical discretion raises significant concerns as the GenAI divination tools cannot decline potentially harmful requests that traditional practitioners would ethically refuse to perform. Moreover, the formation of communities around these tools plays a crucial role in reinforcing beliefs and validating interpretations, with users actively sharing results and seeking feedback, often repeating queries with slight modifications until they receive "satisfactory" results that align with their expectations. Finally, based on these interview findings and observations, we discuss and summarize the typical process by which Chinese users engage with GenAI-based fortune-telling. 

This work is focused on understanding how people \textit{interact} with GenAI in the fortune-telling domain. By analyzing these reported interactions, we find \textit{implications} for how to better design systems applied to spiritual and cultural applications like fortune-telling. Ultimately, our work reveals how AI transforms the \textit{form} of traditional practices while serving the same enduring \textit{goals}.

\section{Background}\label{sec:Background}
\subsection{Chinese Fortune-Telling Practices and Cultural Significance}
\begin{figure}[h]
\centering
\includegraphics[width=1\linewidth]{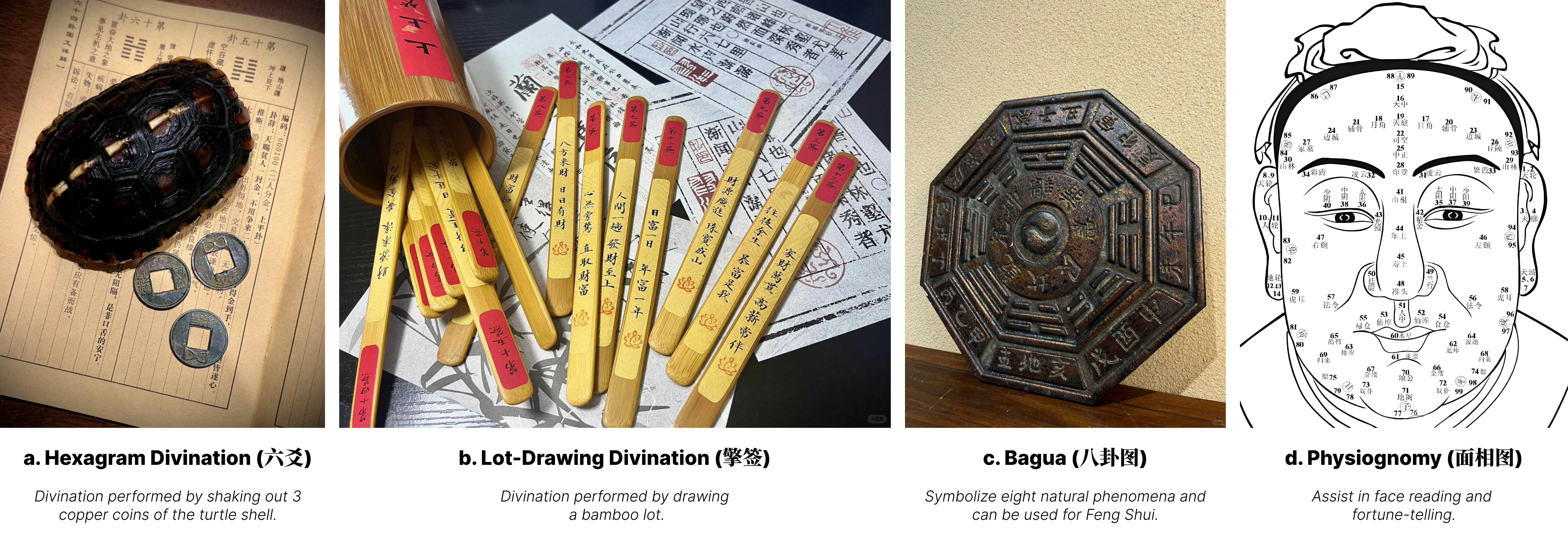}
\caption{Traditional Chinese divination methods examples: (a) Hexagram divination with coins, (b) Lot-Drawing divination with bamboo sticks, (c) \textit{Bagua diagram} for \textit{Feng Shui}, (d) Physiognomy chart for face reading.}
\label{fig: Chinese divination methods examples}
\end{figure}

In Chinese cultural life, divination methods such as \textit{BaZi}, \textit{Ziwei Doushu}, \textit{the I Ching}, \textit{Feng Shui}, and physiognomy serve not merely as a comprehensive worldview but also as practical tools for coping with uncertainty. Rooted in Daoist, Confucian, and folk traditions, these systems draw on the symbolic regimes of heavenly stems and earthly branches, the Five Phases, and hexagrams to analyze destiny, personality traits, interpersonal relationships, and auspicious timing~\cite{Peng2024, schmiedl2023living, Smith1992}. Some examples are shown in Fig.~\ref{fig: Chinese divination methods examples}. These systems embody a distinctly Chinese metaphysical worldview, one that sees fate as shaped through the interaction of Heaven, Earth, and human action. While destiny may be partially predetermined, individuals are encouraged to make informed adjustments based on divinatory insights, reflecting a philosophy that values balance, timing, and harmonious alignment~\cite{schmiedl2023living}. Scholars observe that these practices are as much symbolic and ritualistic as they are predictive, helping individuals impose meaning and order on an uncertain world~\cite{hobson2018psychology, roth1991psychology}.

Framed through the lens of "Distributed Cognition" and "Symbolic Interactionism", traditional divinatory systems can be interpreted as culturally embedded tools that scaffold reasoning, meaning-making, and emotional regulation in the face of uncertainty~\cite{hutchins1995cognition,blumer1986symbolic}. These systems operate by externalizing cognitive processes into shared symbolic artifacts, such as hexagrams, astrological charts, or BaZi matrices, allowing users to interact with and interpret encoded cultural knowledge~\cite{norman1991cognitive,zhang1994representations}. As such, these divinatory logics can be seen as early forms of systematic, symbolic reasoning, positioning fortune-telling as a proto-computational cultural interface long before digital technologies emerged~\cite{dourish2011divining}. This theoretical grounding offers a critical foundation for examining how contemporary GenAI platforms inherit, simulate, or diverge from these longstanding epistemic practices.

Historically, divination was not the exclusive domain of the philosophical elite but pervaded the everyday lives of ordinary people~\cite{roth1991psychology,smith2021fortune}. The spatial arrangements of buildings and interiors according to \textit{Feng Shui}~\cite{Kuo2009}, the reading of facial features in physiognomy~\cite{Cheng2025}, and the use of \textit{BaZi} charts to predict marriage and fertility all reflect a Chinese cosmology in which destiny is at once predetermined and subject to individual influence through ritualized action~\cite{hassin2000facing,roth1991psychology}. 

In recent years, younger generations have re-engaged with traditional divination practices through online fortune-telling, tarot readings, and astrological analysis. Although the mediums have shifted significantly, from religious temples to social media platforms, and from paper-based charts to mobile applications, the underlying cultural motivations and interpretive logics remain remarkably persistent. These practices serve as a means for young people to explore personal identity, alleviate emotional stress, and cultivate a sense of community belonging~\cite{park2024digital}.

Sociological and anthropological studies suggest that this “network fortune-telling” subculture among youth is far from mere superstition; it represents a cultural strategy for resisting anxiety and reshaping self-narratives in a complex society~\cite{lee2024examination}. Individuals here display pronounced cognitive multiplicity, embracing modern scientific rationality alongside enduring metaphysical beliefs~\cite{lee2024examination,lackner2018coping}.

Taken together, these Chinese divination practices constitute more than a body of knowledge; they represent a trans-temporal cultural mechanism that, across historical, social, and psychological dimensions, provides tools for meaning-making, emotional regulation, and identity formation~\cite{roth1991psychology}. The study builds on this rich background to examine how contemporary AI-driven fortune-telling platforms inherit, reconfigure, or contest these deep cultural logics, especially as users engage in algorithmic negotiation, seeking to align machine-generated outputs with longstanding cultural epistemologies like \textit{BaZi}.

\subsection{Why Do People Turn to Fortune-Telling?}

Why do individuals continue to turn to fortune-telling in an era mediated by advanced technology? From a psychological perspective, the Barnum effect offers a critical explanation: individuals tend to perceive vague yet positive statements as highly personal, particularly under the influence of the self-referential effect~\cite{forer1949fallacy, dickson1985barnum}. This phenomenon is especially prevalent in interpretive systems such as astrology and Feng Shui~\cite{meehl1956wanted}. Working in tandem with confirmation bias, individuals selectively absorb information that reinforces their pre-existing beliefs while disregarding contradictory evidence~\cite{fichten1983popular, glick1989fault, nickerson1998confirmation}. These cognitive tendencies, intertwined with emotional needs, drive people to seek psychological certainty in moments of anxiety or ambiguity, making mystical practices like fortune-telling persist as a meaningful form of self-assurance.

However, fortune-telling is not merely a product of individual cognitive bias or illusion. Research in technology and sociology suggests that such practices are deeply embedded in shared cultural logics and community rituals~\cite{song2025walking}. Perceptions of “luck” or “fate” are continuously reinforced through collective narratives, symbolic resonance, and ritualized synchronization, phenomena often conceptualized as collective effervescence~\cite{olaveson2001collective, hobson2018psychology}.

At the social level, traditional religious or spiritual rituals function not only as personal coping mechanisms but as collective sensemaking infrastructures. Ritual synchrony fosters a sense of belonging and “collective effervescence”, which is particularly salient in collectivist cultures~\cite{olaveson2001collective, hobson2018psychology}. Studies have shown that in collectivist societies like China, in-group trust and conformity pressures are 27\% stronger than in individualist cultures~\cite{norenzayan2016cultural}. In collectivist societies like China, such trust-building mechanisms are particularly pronounced~\cite{norenzayan2016cultural}. As digital media enters this space, from spiritual apps and live-streamed divinations to GenAI-powered fortune-telling systems, technological affordances increasingly shape how mystical beliefs are performed and perceived~\cite{hill2022roar, byrne2022spooky}. In this context, fortune-telling is no longer just a form of emotional regulation, but a socio-cultural mechanism through which uncertainty is collectively negotiated, and meaning is co-constructed via human–technology interaction.

At the level of cultural and cognitive processing, the temporary suppression of analytical thinking is a crucial precondition for the formation of mystical beliefs~\cite{gervais2012analytic, bronstein2019belief}. Gervais et al. and Bronstein et al. suggest that analytical suppression enhances intuitive receptivity to supernatural ideas, while Pennycook et al. further demonstrate that under cognitive load, individuals’ acceptance of mystical explanations increases by 38\%~\cite{pennycook2012analytic}. These findings help explain behavioral patterns such as users turning to AI fortune-telling apps late at night when decision fatigue is high and intuitive reasoning (System 1) dominates~\cite{de2008conflict, evans2013dual, nagar2012you}. This intuitive reliance also opens space for affective heuristics and self-serving interpretations, as noted in dual-process theories of cognition~\cite{baumard2013religious}.

However, GenAI systems challenge this structure by becoming both technological actors and cultural intermediaries. On one hand, large language models carry epistemologies embedded in their training data, which are often dominated by Western norms and rationalities~\cite{tao2024cultural, li2024culturepark}. This creates a potential mismatch with non-Western spiritual frameworks and may result in cultural erasure or distortion. On the other hand, GenAI tools also offer novel modes of trust construction, where algorithmic opacity is not necessarily a drawback, but becomes part of the mystical allure~\cite{bender2021dangers, noble2018algorithms}. The “stochastic parrot” critique reminds us that these tools may simulate understanding without semantic depth, yet users continue to engage with them in ways that are emotionally resonant and culturally adaptive~\cite{bender2021dangers,noble2018algorithms}.

In this sense, explainability becomes not just a technical challenge, but a cultural design problem~\cite{peters2024cultural}. Work by Peters et al. reveals that user trust in AI explanations depends heavily on the alignment between system behaviors and culturally familiar metaphors of agency. Thus, GenAI fortune-telling interfaces that incorporate zodiac, I Ching, or Tarot imagery are not trivial; they function as bridges between algorithmic logic and symbolic meaning.

From a contemporary perspective, with the rise of GenAI-driven fortune-telling tools (such as tarot robots, horoscope-based GPTs, or personality analysis AIs), the concept of “superstition” is being profoundly reshaped. Recent studies show that users increasingly perceive these tools as forms of digital emotional companionship and devices for emotional exploration, rather than traditional “prediction mechanisms”~\cite{banerjee2024exploring}.

As Banerjee et al. point out, these AI-powered divination practices blur the boundaries between the rational and the irrational, repackaging previously stigmatized behaviors as legitimate forms of self-understanding~\cite{banerjee2024exploring}. Rather than eliminating superstition, AI recontextualizes it through its customizable, depersonalized, and instant interaction format, freeing it from a rigid binary of belief vs. disbelief and turning it into a more flexible emotional practice~\cite{carpentierreligious}. Notably, many users are aware of the computational logic behind these tools, yet still project genuine emotional meaning onto the interaction~\cite{da2022cyber,muir2021hci}. This willingness to “play along” despite knowing the artificiality reflects a kind of postmodern gamified belief, a conscious “performance of faith” through which users seek psychological comfort, social connection, and identity construction.

Therefore, the reconstruction of superstition through GenAI is not a simple process of “technology replacing tradition.” Instead, it reflects a deeper negotiation between cultural, cognitive, and social mechanisms. In this process, belief becomes a designed interactive structure, and superstition transforms into a new kind of socio-psychological infrastructure, supported by intelligent systems. This raises new questions for HCI research: should we design belief into AI systems, or at least understand how people assign faith-like meaning in their interactions with AI?

\subsection{Algorithmic Divination with Cultural Alignment, Participation, and System Design}

Recent advances in GenAI-powered divination systems reveal three interconnected dimensions of inquiry. Technically, researchers have explored diverse algorithmic implementations to bridge data-driven prediction with spiritual symbolism. McNutt~\cite{Mcnutt2020DiviningIV} developed Sortilège, a visual analytics system that maps algorithmic outputs onto tarot metaphors, while Pichlmair’s procedural generation tools~\cite{Pichlmair2020ProceduralGF} semi-automate tarot deck creation through rule-based algorithms requiring human semantic refinement. Narrative generation models, such as Sullivan’s tarot-to-story system~\cite{Sullivan2018TarotbasedNG}, leverage NLP to transform divinatory symbols into cinematic predictions, whereas Shin’s cross-modal framework~\cite{Shin2024FromPT} translates academic concepts into tarot-inspired design cards. Lustig’s machine learning experiments~\cite{Lustig2022FromET} further demonstrate how training models on historical tarot datasets enables personalized deck generation. These technical approaches collectively grapple with balancing algorithmic pattern recognition with human interpretative frameworks.

Psychologically, users engage with AI divination through complex belief dynamics. Lee’s studies~\cite{Lee2024SuperintelligenceOS, Lee2024ThePO} identify “rational superstition,” where users validate AI predictions via pre-existing astrological beliefs, selectively accepting positive outcomes as personalized insights. Shein’s work~\cite{Shein2014RelationshipBS} highlights cognitive polyphasia, users’ simultaneous adherence to scientific rationality and spiritual practices, necessitating ethical safeguards in prediction delivery. Historical analyses by Pooley~\cite{Pooley2023PaperTF} and Tost~\cite{tost2024companion} contextualize card-based divination as enduring tools for emotional processing, a role modern systems like Tarotoo~\cite{Lustig2022FromET} enhance through iterative feedback loops. This creates a paradox: users demand AI’s perceived objectivity while insisting on cultural-psychological alignment of outputs.

Contextually, tarot and card-based systems permeate diverse domains. In design, Chung’s interaction cards~\cite{Chung2015InteractionTA} and Michelson’s world-building frameworks~\cite{Michelson2024WorldingWT} repurpose tarot for creative ideation, while Skovbjerg’s gamified system~\cite{Skovbjerg2022DevelopingPT} adapts divinatory logic for teacher training. Artistic reinterpretations, such as Sarah’s batik adaptations~\cite{Sarah2020REPRESENTASIGP} and Short’s installations~\cite{short2016card}, aestheticize spiritual symbolism. AI-augmented tools like Ntelia’s Manifesting~\cite{ntelia2022manifesting} and Pichlmair’s procedural generators~\cite{Pichlmair2020ProceduralGF} hybridize traditional divination with computational creativity, enabling rapid prototyping of culturally-grounded spiritual interfaces.

Some classic works~\cite{grudin1988cscw, dourish1992awareness, resnick1994grouplens} warn that design failures often stem from a mismatch between technical mechanisms and actual user needs, particularly when a system fails to align with users' cultural expectations and social contexts. For GenAI fortune-telling tools, algorithmic accuracy alone is insufficient; their interactive flows, feedback loops~\cite{dourish1992awareness}, and recommendation architectures~\cite{resnick1994grouplens} must embed cultural and spiritual symbols familiar to users, or they will face significant adoption barriers.

Two critical gaps emerge from this body of work. First, existing research predominantly focuses on Western divination traditions (e.g., Tarot, astrology) and their HCI applications~\cite{Shein2014RelationshipBS, Michelson2024WorldingWT}, while non-Western spiritual practices, despite their distinct epistemological foundations and societal influence, remain underexplored. This omission is particularly significant given that both GenAI and many traditional divination systems operate through data pattern recognition, yet their cultural legitimacy and user acceptance mechanisms may vary drastically. Second, the perceptual landscape of GenAI divination users remains poorly mapped. While studies address belief dynamics~\cite{Lee2024SuperintelligenceOS} and cognitive polyphasia~\cite{Shein2014RelationshipBS}, there is limited empirical understanding of 1) how different demographics reconcile AI's data-driven predictions with pre-existing spiritual beliefs, and 2) whether users perceive GenAI as a continuation of traditional divination logic or a distinct technological paradigm. This gap is compounded by the lack of comparative studies analyzing how cultural context mediates trust in AI-generated spiritual guidance, especially in regions like China where divination practices are deeply intertwined with local cosmologies. 

This study aims to address critical gaps in existing literature, particularly the lack of non-Western perspectives and limited understanding of user perception, through field interviews and cultural analysis. We systematically explore how Chinese users engage with GenAI-powered fortune-telling tools in practice, how they evaluate these tools’ accuracy and effectiveness based on personal experience, and how they construct trust and meaning through community interactions. By linking algorithmic generation mechanisms with users’ cultural expectations, cognitive judgments, and social contexts, this study reveals how GenAI fortune-telling is emerging in the Chinese context as a novel form of “algorithmic cultural practice” that negotiates between technology, belief, and emotion.

\section{Methods}\label{sec:Methods}
Semi-structured interviews were conducted with 22 participants through Tencent Meeting, WeChat, Zoom, and in-person sessions. The objective was to explore how individuals engage with GenAI fortune-telling tools and the interpretive or spiritual meanings they derive from these interactions. We also conducted a three-week digital ethnography with 1,842 posts. This involved observing and analyzing user interactions, community discussions, and platform features on popular GenAI divination platforms, such as Cece and related WeChat groups and forums. All research procedures were reviewed and approved by the institutional ethics committee, with strict adherence to protocols for informed consent and data privacy. 

\subsection{Participants and Recruitment}

We conducted semi-structured interviews with 22 participants (14 female, 8 male) aged 18-52 with an average age of 27.2 (see Table.\ref{tab:my_label}). Participants were recruited through direct outreach and public social media posts on WeChat and RedNote. Eligibility was based on having used at least one GenAI tool or platform for divination. The final sample included individuals from diverse professional backgrounds, including industrial design (n=5), art (n=4), technology-related fields (n=3), and others such as law, accounting, and early childhood studies. Their experience spanned across different GenAI platforms, from general LLMs like ChatGPT and DeepSeek to specialized GenAI fortune-telling platforms, with popularity using Cece for divination. Only 4 participants reported experience with specialized GenAI fortune-telling tools (Ziwei Doushu GenAI, Quin GenAI, TianGong AI, etc.), and no experience with LLMs. These tools were used for various divination methods, including astrology, \textit{BaZi}, tarot, face reading, and \textit{Feng Shui}. To distinguish between habitual reliance and casual curiosity, we investigated participants' usage frequency based on their engagement over the past six months: Frequent users (n=6) interacted with GenAI fortune-telling tools at least once a week; Occasional users (n=13) used the tools 1–3 times per month; and One-time users (n=3) had only a single try.

While our sample captures diverse perspectives in terms of age, gender, educational and professional background, it has some limitations. Only six participants were frequent users, which may constrain understanding of long-term engagement patterns. The sample was also skewed toward younger adults and female participants, reflecting, to some extent, the demographic tendencies of GenAI fortune-telling adoption in China. Consequently, experiences of older adults or infrequent users may be underrepresented.

Throughout the study, ethical integrity was maintained, with informed consent obtained from all participants and strict protocols followed as per institutional Internal Review Board (IRB) requirements. Participants were compensated with 50 CNY for their time and contribution to the study.

\begin{table}[h]
    \vspace{-0.2cm}
    \caption{Demographic Information of Participants}
    \centering
    \renewcommand{\arraystretch}{1.5}
    \resizebox{\textwidth}{!}{
    \begin{tabular}{ccccccc} 
        \hline 
        \textbf{ID} & \textbf{Gender} & \textbf{Age} & \textbf{Interview Mode} & \textbf{GenAI Platforms (Fortune-telling Methods)} & \textbf{Usage Freq.} & \textbf{Major and Domain} \\
        \hline 
        P1 & Female & 24 & Online & Cece (Astrology, BaZi) & Frequent & Law\\
        P2 & Female & 24 & In-person & Cece (Tarot) & Frequent & Art\\
        P3 & Female & 24 & In-person & ChatGPT (Dream Analysis) & Occasional & Industrial Design\\
        P4 & Male & 22 & Online & ChatGPT (Astrology, BaZi, Face \& Palm, Tarot) & Once & Industrial Design\\
        P5 & Male & 22 & In-person & ChatGPT (Astrology) & Once & Industrial Design\\
        P6 & Female & 29 & Online & Cece (Astrology, Tarot) & Frequent & Industrial Design\\
        P7 & Female & 24 & Online & ChatGPT (Tarot) & Occasional & Art\\
        P8 & Male & 22 & Online & ChatGPT (BaZi) & Once & Industrial Design\\
        P9 & Female & 23 & Online & ChatGPT (BaZi) & Occasional & Art\\
        P10 & Female & 22 & Online & WenZhen (BaZi) & Occasional & Accounting\\
        P11 & Female & 24 & In-person & ChatGPT (Astrology, Face \& Palm, Tarot) & Occasional & Architecture\\
        P12 & Female & 26 & Online & Cece (Tarot, Liu Yao); ChatGPT (Tarot, Liu Yao) & Frequent & Administrative Mgmt.\\
        P13 & Female & 30 & Online & Cece (BaZi, Tarot, Liu Yao); WenMoTianJi (Astrology) & Frequent & Early Childhood Studies\\
        P14 & Male & 18 & Online & TianGong AI (Astrology, BaZi, Face \& Palm, Feng Shui) & Occasional & Art\\
        P15 & Female & 27 & Online & ChatGPT (BaZi); Cece (Astrology) & Occasional & Computer Info. Mgmt.\\
        P16 & Male & 21 & Online & Ziwei Doushu GenAI (Astrology); FateTell (Feng Shui); Quin (Tarot) & Occasional & Architecture\\
        P17 & Female & 24 & Online & Cece (Astrology) & Occasional & Computer Science\\
        P18 & Female & 26 & Online & Cece (Tarot); Ernie Bot (Tarot) & Occasional & Electronic Info. Eng.\\
        P19 & Female & 32 & Online & Cece (BaZi); Deepseek (BaZi) & Frequent & Education \\
        P20 & Male & 34 & Online & Deepseek (BaZi); Ernie Bot (BaZi) & Occasional & Product Mgmt. \\
        P21 & Male & 48 & Online & Deepseek (BaZi) & Occasional & Architecture \\
        P22 & Male & 52 & Online & WeChat Mini Programs (BaZi, Palmistry); Deepseek (BaZi) & Occasional & Business Mgmt. \\
        \hline 
    \end{tabular}
    }
    \label{tab:my_label}
    \vspace{-0.2cm}
\end{table}
  
\subsection{Interview Procedure}
We designed semi-structured interviews on Tencent Meeting, WeChat, Zoom, and in-person sessions, lasting between 30 to 60 minutes. The choice of interview mode was participant-driven to ensure comfort and convenience. We conducted 4 in-person interviews (P2, P3, P5, P11), while the remainder were online via Tencent Meeting (n=12), WeChat (n=4), and Zoom (n=2). No significant discrepancy in narrative depth was observed across modes.

\begin{figure}[h]
\vspace{-0.2cm}
    \centering
    \includegraphics[width=1\linewidth]{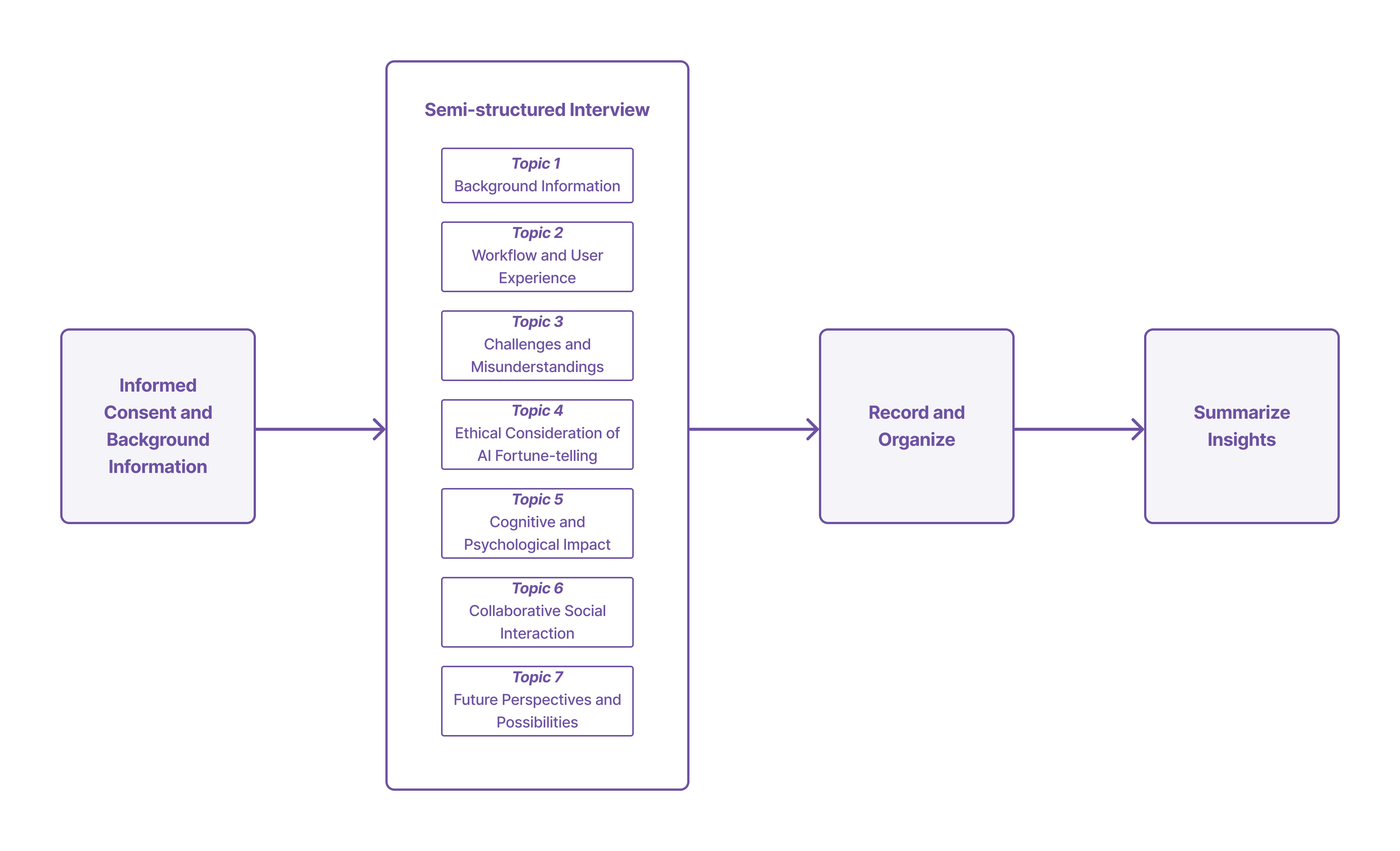}
    \vspace{-0.8cm}
    \caption{Interview Flowchart: We conducted semi-structured interviews with seven topics (Background Information, Workflow and User Experience, Challenges and Misunderstandings, Ethical Consideration of GenAI Fortune-telling, Cognitive and Psychological Impact, Collaborative Social Interaction, and Future Perspectives and Possibilities).}
    \label{fig: Interview Procedure}
    \vspace{-0.2cm}
\end{figure}

Participants were informed beforehand that the entire discussion would be recorded and transcribed. Following a structured protocol, we first collected background information about participants' experience with various GenAI fortune-telling applications, their usage patterns, and the reasons they started using GenAI fortune-telling. During the interviews, the researcher guided participants to reflect on their workflow and user experience with GenAI fortune-telling tools, followed by discussions about challenges and misunderstandings they encountered. The researcher then encouraged the participants to discuss the ethical considerations of GenAI fortune-telling and its cognitive and psychological impact on the participants. For example, the researcher might ask participants how they interpreted and acted upon GenAI-generated predictions. Additionally, participants were encouraged to share their collaborative social interactions around GenAI fortune-telling, such as how they shared and discussed predictions with others. The interviews concluded with discussions about future perspectives and possibilities for GenAI fortune-telling (see Fig.~\ref{fig: Interview Procedure}). All interviews were conducted in Chinese, and the recordings were transcribed via online meeting rooms and translated into English by the research team for subsequent analysis, with all personal identifiers removed during the process.

The detailed semi-structured interview outline is provided in Appendix. \ref{app: questions}, which includes discussions on both positive and negative perceptions of AI fortune-telling, also encouraged to share participants' experiences regarding negative predictions.

\subsection{Digital Ethnography Procedure}
In addition to conducting interviews, we employed a three-week digital ethnography (from March 4 to March 25, 2025) to gain deeper insight into how users interact with GenAI-based fortune-telling tools in real-world contexts. This approach involved systematically observing user behaviors, discourses, and cultural practices across multiple online platforms where such interactions are frequently shared and discussed. Based on insights from the interviews, we identified WeChat, RedNote, and user communities embedded within GenAI apps like Cece as the most frequently mentioned spaces for discussion and engagement. During the observation period, we focused on both public-facing posts and semi-public interactions such as comment threads, discussion forums, and experience-sharing notes, with particular attention paid to how users describe, interpret, and evaluate their divination experiences. To ensure analytical consistency, we applied inclusion and exclusion criteria when selecting posts. Posts were included if they documented users’ direct engagement with GenAI fortune-telling tools—such as sharing AI-generated readings, reflecting on their accuracy, or narrating emotional responses. General horoscopes (e.g., “Taurus weekly forecast”) that were non-interactive, not AI-generated, or lacked evidence of personal engagement were excluded. Similarly, purely promotional content, reposted memes, or materials unrelated to divination practices were omitted. This allowed us to focus on data that demonstrated meaningful user-tool interaction, interpretation, and community discussion.

Specifically, we manually archived a total of 1,842 posts and interactions, including 803 RedNote notes, 112 WeChat articles, 589 user-generated posts from Cece community forums, and 338 comment threads and replies extracted from in-app discussion features. All collected content was documented through anonymized screenshots, totaling over 450 visual records, which were analyzed to identify recurring themes, metaphors, affective responses, and platform-specific discursive norms. These records served to complement the interview data by offering rich ecological context and highlighting how GenAI fortune-telling is embedded in users’ everyday social and emotional lives. 

Given the public or semi-public nature of the platforms observed, we did not perceive significant ethical risks in conducting online participant observation. To ensure the protection of user privacy, all personally identifiable information (including usernames, avatars, and chat handles) was removed, and only anonymized excerpts were used in analysis. All procedures adhered to institutionally approved ethical guidelines, and no identifiable data were retained.

\subsection{Data Analysis}
We employed a combination of open coding, thematic analysis~\cite{clarke2017thematic}, and network ethnographic methods~\cite{kozinets2015netnography} to interpret the interview transcripts, following principles of grounded theory~\cite{glaser2017discovery}. Open coding and thematic analysis helped organize recurring patterns in participants’ narratives into coherent themes across cases, and network ethnography provided a lens to situate individual experiences within the wider socio-technical and cultural ecologies of GenAI divination.

In the first step, three researchers independently coded the interview transcripts and screenshots of social media posts to ensure analytical rigor and minimize bias. Initial codes were derived inductively from participants' descriptions of their experiences with GenAI-based fortune-telling tools and online collected data about comment threads, discussion forums, and experience-sharing notes.

Subsequently, the findings were organized into seven main themes that aligned with the semi-structured interview framework: (1) Background Information, examining participants' motivations and initial encounters with GenAI fortune-telling; (2) Workflow and User Experience, analyzing how participants interacted with different platforms; (3) Challenges and Misunderstandings, identifying common issues users faced; (4) Ethical Consideration of GenAI Fortune-telling, exploring participants' moral perspectives; (5) Cognitive and Psychological Impact, examining how GenAI predictions influenced users' thinking and emotions; (6) Collaborative Social Interaction, investigating how users shared and discussed AI-generated predictions; and (7) Future Perspectives and Possibilities, gathering user insights on potential developments. 

Following this, the team engaged in reflexive dialogue and used iterative discussions to identify and refine themes within the data, analyzed in conjunction with the ethnographic findings. This process facilitated the categorization of the interview data into three core topics: "Perceptions of Effectiveness and Trust", "Social Interactions in Online Community", and "Common GenAI Fortune-telling Process". These themes were then collaboratively reviewed and further refined by the full research team to ensure their consistency, relevance, and alignment with both the interview content and the broader ethnographic context. 

The refinement process moved from descriptive initial codes to analytical themes. For example, initial codes such as "sharing screenshots on WeChat" were grouped under the intermediate category of "social sense-making," which eventually evolved into the final core theme: "Collective Validation in Online Communities." Similarly, codes regarding "AI's lack of moral judgment" in someone's interview record were refined into "Ethical Considerations" of the 7 themes.

\subsection{Positionality}
The research team consists of HCI researchers with backgrounds in social psychology and design. Three researchers are Chinese speakers, providing the team with the ability to understand linguistic nuances and cultural metaphors (e.g., BaZi concepts) within the data. The team members also have previous experiences in academical qualitative analysis to maintain a neutral stance. We acknowledged our prior familiarity with GenAI tools as a potential bias which may influence data interpretation; we addressed this through reflexive memoing and cross-coder triangulation during data analysis.

\section{Results}\label{sec:Results}
We examined the utilization, perception, and communal influence of GenAI-based fortune-telling among Chinese users. The findings illuminate how users navigate these tools to meet their personal and cultural expectations in fortune reading, their perceptions of the effectiveness and accuracy of AI-generated divinations, and the role of GenAI divination within their communities, especially in shaping trust and social dynamics. Based on the findings from the interviews and network ethnography, we summarized the process of GenAI fortune-telling.

\subsection{Perceptions of Effectiveness and Trust}

\subsubsection{Motivation for fortune-telling is the key factor influencing acceptance.}
\label{sec:4.1.1}

Users' motivations for engaging with GenAI divination play a significant role in shaping their level of acceptance and response to the results. Five different motivations lead to varied attitudes toward the interpretations provided by the GenAI tool, influencing the depth of users' engagement and the likelihood of their reliance on these divination results.

\begin{figure}[h]
\centering
\includegraphics[width=1\linewidth]{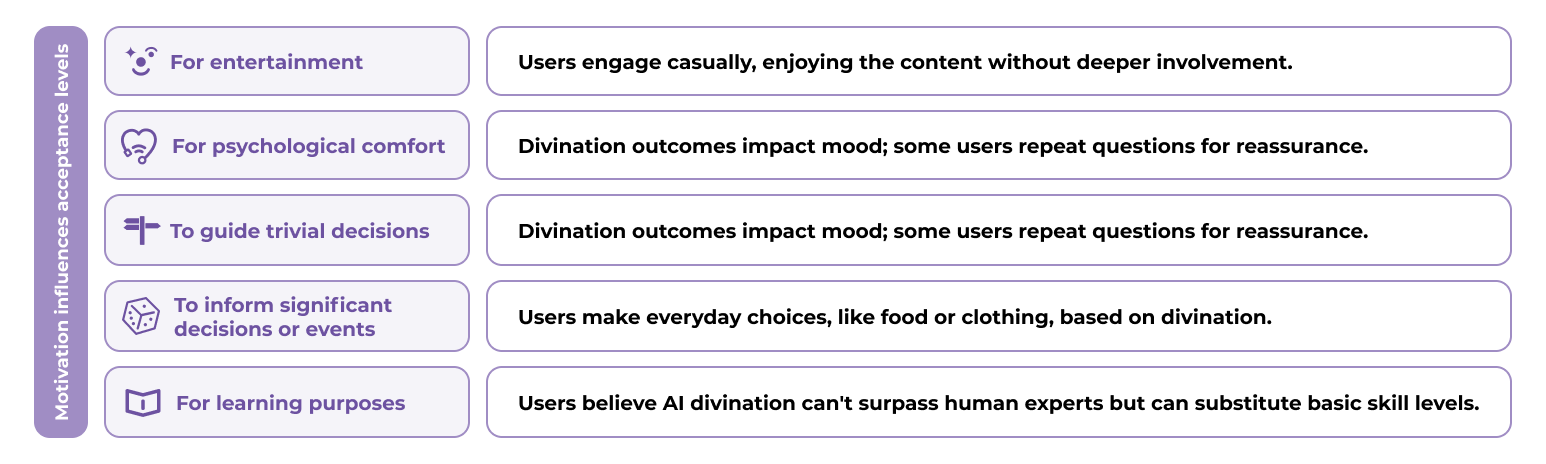}
\vspace{-0.6cm}
\caption{Under the five different divination motives, the influences received by users and their acceptance levels vary.}
\label{fig: Theme 1}
\vspace{-0.2cm}
\end{figure}

Some users approach GenAI divination purely as a form of lighthearted \textbf{entertainment}. For instance, one participant noted, \textit{"It's just a fun activity; I don't take it too seriously. I often use it after making decisions rather than before"} (P2). They maintain a superficial attitude toward the results, engaging with the divination more as casual amusement.

For others, GenAI divination serves as a source of \textbf{psychological comfort}. These users seek emotional reassurance rather than strict accuracy. For instance, one participant shared that they \textit{"use it for emotional release, like venting to the AI when feeling frustrated"} (P16), while another mentioned, \textit{"It gives me peace of mind, not necessarily an accurate answer, but a quick way to ease my mind"} (P3). This motivation drives some users to repeatedly engage with the GenAI, subtly adjusting their questions each time, aiming to receive a response that offers the reassurance or encouragement they seek. As P22, a business owner and manager, said, “...When I’m not satisfied with the divination result, I just make it tell again and again until I’m happy with it.”

In addition, another major theme for users to use GenAI divination is to assist in \textbf{decision-making}, and the size of the matter to be decided also brings about a huge difference in attitude. Some users utilize GenAI divination for minor, everyday choices. One participant mentioned, \textit{"I sometimes let it decide small things, like the 'lucky food' of the day"} (P12). In this context, the results serve as playful routine decisions, adding an element of fun to daily life without deeply affecting major decision-making. Another participant shared, \textit{"I find it helpful for quick, low-stakes guidance. It adds a bit of amusement to small decisions like what to wear"} (P16). 

For decision-making purposes, the other group of people in a smaller size turns to GenAI divination for guidance on major life events or significant decisions. Although they seek insights, they often remain cautious, with one participant noting, \textit{"AI isn't something I'd rely on for big decisions, but it can offer a different perspective"} (P14). Interestingly, some users return to GenAI after a decision has been made, using it to reinforce and validate their choices. In this post-decision phase, they seek confirmation from GenAI results to reassure them of their chosen path. Participant P14 elaborated on this practice, sharing that she often consults GenAI to \textit{"reconfirm decisions whenever doubt arises, especially in high-stakes situations"}. This post-decision use of GenAI divination reflects a desire for responses that align with their existing perspective, thereby strengthening their confidence and sense of certainty in the choices they've made.

Finally, some users engaged with GenAI divination primarily out of a desire to \textbf{learn and compare} it with traditional practices. These individuals tended to approach the tool with a more critical lens. As one participant put it, \textit{"I'm interested in understanding the differences, but I find AI lacks the depth of a real human fortune-teller"} (P9). These individuals are often more skeptical of the AI's capabilities, perceiving it as a limited learning tool. Another remarked, \textit{"It's insightful to some extent, but it feels like AI can't capture the same intuition that humans do"} (P3). Their acceptance level remains relatively low, as they view GenAI divination as a supplementary rather than a fully reliable guide.

Through these varying motivations, we observe that a user's specific emotional state, whether driven by playful curiosity or deep-seated anxiety, dictates the ontological status of the AI agent. For users seeking entertainment (e.g., P2), the interaction is a low-stakes social game, often shared with peers to foster connection, functioning as a "social lubricant" for extroverted engagement. Conversely, for users driven by a need for psychological comfort or navigation through desperation (e.g., P3, P16), the AI assumes the role of a private, non-judgmental confidant. This is particularly significant for individuals who may experience loneliness or possess introverted traits; for them, the AI provides a "safe space" to externalize inner conflicts without the social pressure inherent in human interaction. Thus, the reasoning and relationship processes are starkly different: the former collaborates with AI to amplify joy and social bonding, while the latter collaborates with AI to reconstruct a fragile sense of self and emotional stability. This nuance highlights that GenAI fortune-telling is not a monolithic practice but a situated interaction deeply entangled with users' individual psychological well-being and their broader social positioning.

\subsubsection{GenAI divination results are readily available but overly standardized and without mystery.}

Users hold contrasting views on the results generated by GenAI divination tools. While many appreciate the ease and speed of access, others express concerns about the lack of personalization and emotional depth. This subsection unpacks these different perspectives, highlighting how convenience often comes with a trade-off in perceived authenticity or mystical value. Fig.~\ref{fig: Theme 1} summarizes the key positive and negative perceptions identified through our interviews and online observations, offering an overview of how users interpret GenAI-generated predictions.

\begin{figure}[h]
\centering
\includegraphics[width=1\linewidth]{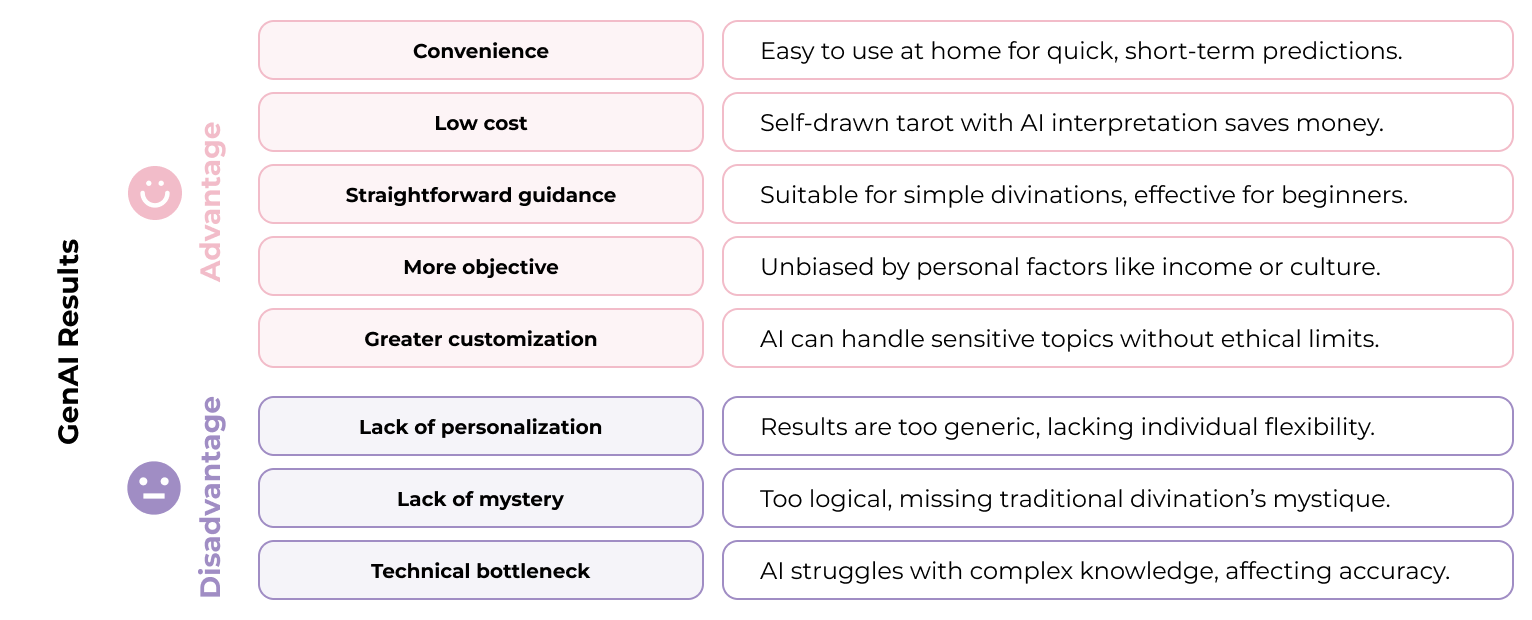}
\vspace{-0.5cm}
\caption{Users' views on the advantages and disadvantages of GenAI divination results.}
\label{fig: Theme 1}
\vspace{-0.3cm}
\end{figure}

One key advantage of GenAI divination is its \textbf{convenience}. Many users appreciate the accessibility it offers, allowing them to access daily or short-term forecasts at home without scheduling an in-person consultation. For instance, participant P1 notes, \textit{"It's faster and more cost-effective compared to offline services, especially for things like Tarot, which I typically consult when I'm under stress or facing life issues"} (P1). Second, the \textbf{low cost} of GenAI platforms also makes them attractive. By facilitating self-drawn tarot readings and AI-generated interpretations, users can engage with divination without incurring the expense of professional fortune-tellers. As P2 mentions, \textit{"AI fortune-telling allows me to save money, especially when I need reassurance without spending a lot on traditional services"} (P2). Third, GenAI has proven especially praiseworthy for users seeking simple divination results, as it effectively meets their needs for \textbf{straightforward guidance}. For instance, P9 noted that AI provides \textit{"satisfactory results for simple questions and everyday guidance, enough to replace a basic level of human interpretation"} (P9). Fourth, GenAI's \textbf{objectivity} appeals to users who prefer unbiased analysis free from external influences. Participant P5 noted, \textit{"AI's approach felt more neutral and objective...a human practitioner might try to appeal to what clients want to hear"} (P5). Similarly, P8 described this neutrality as beneficial, noting that it \textit{"makes the communication feel more genuine because it isn't influenced by my own expectations"} (P8). In addition, GenAI enables \textbf{greater customization}, allowing users to select specific topics they want to explore, rather than being limited to the subjects a human diviner might prioritize. This flexibility includes a \textbf{wider range of topics}, even those that traditional diviners might avoid due to ethical concerns. Participant P11 pointed out that AI does not hesitate to address questions related to privacy or life-and-death matters—areas where human diviners often exercise caution. They noted, \textit{"a fortune-teller with professional ethics might refuse certain questions, but AI seems unrestricted"} (P11), highlighting the extensive and customizable scope AI brings to divination.

However, participants also pointed out disadvantages, particularly around GenAI's \textbf{lack of personalization}. For example, P5 notes, \textit{"AI lacks the human touch that makes traditional readings unique to the individual. The responses feel like they come from a preset model, not tailored to me"} (P5). P14 further criticized the rigid nature of AI interpretations, stating, \textit{"The results often seem too templated and miss the nuance that an experienced fortune-teller might capture"} (P14).

Another common criticism is the \textbf{conformity to expectations} and \textbf{lack of mystery} inherent in GenAI readings, which can make the experience feel too logical or formulaic. P6 remarked, \textit{"Fortune-telling with AI doesn't have that mysterious aura—it's missing the spiritual ambiance that I associate with divination"} (P6). P11 echoed this sentiment, adding, \textit{"When AI gives me a reading, it feels flat and unexciting. There's no ritual or mystique, which is part of what I enjoy about traditional methods"} (P11).

In addition to this lack of ritual, some participants noted that GenAI divination responses tend to conform to general positivity and avoid unsettling content, leading to a lack of depth. As P19 commented, "The results calculated by GenAI are too 'normal'. It always says 'stay hopeful', 'you are worthy of love'...which is of course good, but it doesn't really arouse my real emotions." This sense of predictability diminished their emotional engagement with the tool. As a result, the divination feels less meaningful, even if it's emotionally comforting.

Finally, participants cited a \textbf{technical bottleneck} in GenAI's capacity to interpret complex or highly personalized divination. P8 shared an experience, saying, \textit{"ChatGPT can't handle intricate \textit{BaZi} calculations accurately; it lacks the depth needed for multi-layered interpretations"} (P8). P13 agreed, adding, \textit{"When it comes to nuanced, contextual insights, AI just doesn't compare to a seasoned fortune-teller who can read between the lines"} (P13).

\subsubsection{Compared to traditional fortune-telling, GenAI feels safer to users, yet ethical concerns remain.}

For many Chinese users, GenAI divination seems to alleviate some privacy concerns typically associated with traditional human diviners. Users often feel more at ease providing personal information to AI systems, likely due to the perceived anonymity and reduced need to directly share sensitive details with a human practitioner. For instance, participant P8 noted, \textit{"I'm not too concerned about birth dates, but when it comes to sharing personal thoughts or concerns, privacy issues might arise (with human fortune-tellers)"} (P8). Similarly, participant P18 observed that using GenAI divination poses relatively low privacy risks, commenting, \textit{"in daily life, privacy is already compromised in so many ways, one more instance doesn't make a big difference"} (P18). 

However, participants remain cautious, particularly in cases involving deeper personal details. While some feel they can safely use GenAI for minor inquiries, such as daily guidance, they avoid it for more complex issues. As P2 remarked, \textit{"I use AI for smaller queries and when I want quick results, but for something more personal or emotional, I still prefer face-to-face sessions"} (P2). 

Importantly, participant P10 emphasized a critical ethical gap in GenAI divination: AI lacks the discernment to reject requests that may infringe on others' privacy. Unlike traditional diviners, who are often selective about the types of questions they answer, particularly on sensitive issues like privacy, health, or life-and-death matters. GenAI divination lacks similar ethical constraints but enlarges the scope of divination. Participant P11 noted this concern, explaining, \textit{"A fortune-teller with professional ethics might refuse certain questions, like those about life and death or others' privacy, but AI seems unrestricted"} (P11). Similarly, P17 commented on the boundaries human diviners set, such as refusing readings for younger users or those with mental health issues, adding, \textit{"Some older fortune-tellers won't read for people under 20, but AI doesn't have these ethical considerations"} (P17).

This lack of ethical constraints leads to potential risks in high-stakes domains. As P10 illustrated with an example, a user might ask, \textit{"Should I agree to a doctor performing surgery on my child to save his life? What is the success rate?"} A human fortune-teller would likely refuse such a sensitive question due to its serious implications, whereas GenAI might simply offer comforting but uninformed responses. Similarly, in relationship contexts, we observed that fortune-telling communities involve users asking AI to judge moral correctness or predict breakups. Common questions include, "How long will this relationship last?" or "Can I get along harmoniously with a Scorpio boy?" Responses to such questions are highly impactful; while most users may not strictly follow GenAI's advice, the answers can still leave negative psychological impressions, potentially leading to irreparable harm in their relationships.

Therefore, without a framework for responsible refusal, GenAI risks providing guidance that lacks the necessary empathy and caution.

\subsubsection{Chinese GenAI models elevate cultural specificity in divination, balancing versatility and ritual.}

Following the rise of interest in GenAI-based fortune telling, we observed a notable boom in the use of DeepSeek for divination purposes. According to our online ethnographic observations, after DeepSeek's popularity surged, more than 2 million posts on China’s largest social platform, WeChat, mentioned "DeepSeek Divination" in February 2025 alone. This surge reflects not only the accessibility of GenAI tools but also users’ enthusiasm for culturally resonant applications. Two key findings emerge from this trend.

\begin{figure}[h]
\centering
\includegraphics[width=1\linewidth]{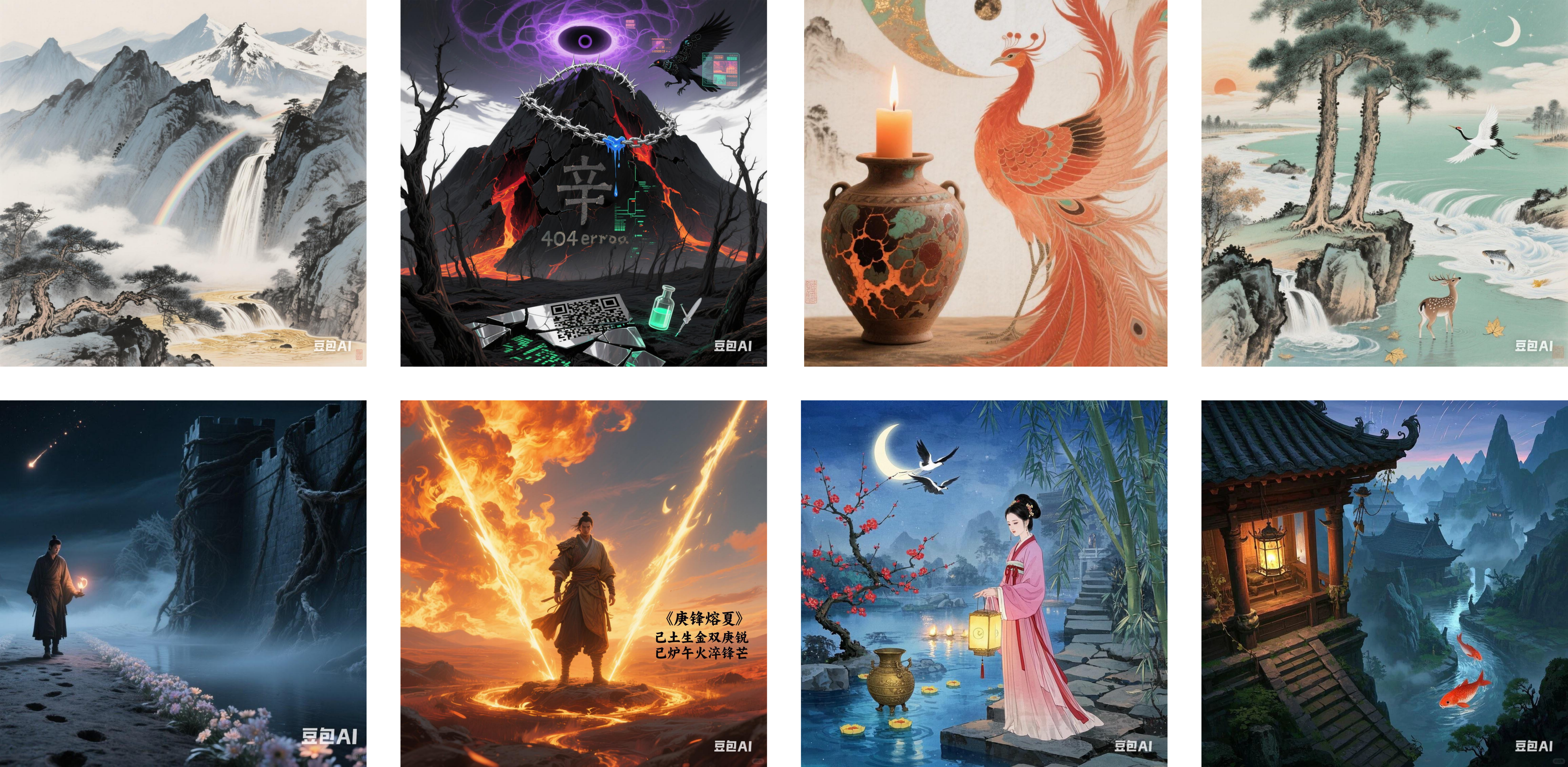}
\caption{Doubao AI can generate visual divination results based solely on \textit{BaZi} of one's birth chart.}
\label{fig: doubao}
\end{figure}

Chinese GenAI models show distinct advantages in supporting traditional Chinese divination practices. These models are better attuned to the symbolic systems, linguistic nuances, and cultural logic embedded in Chinese metaphysics. For example, Participant P21 highlighted that \textit{"tools like DeepSeek can generate more detailed and culturally accurate interpretations for systems like \textit{Ziwei Doushu}, which involves intricate symbolic elements that require contextual understanding."} Participant P20, a new-generation parent, emphasized the value of Chinese-developed GenAI for generating meaningful baby names rooted in the \textit{Five Elements} and tonal harmony purpose. He said that \textit{"We wanted something special, but still meaningful. AI gave us options with good explanations"}. He also emphasized his preference for Chinese-developed AI tools, believing they better capture the cultural nuances embedded in Chinese naming traditions, with non-localized models like ChatGPT falling short. \textit{"Things like character meanings, the Five Elements, or tonal harmony—foreign tools just don't get them. Even though overseas AI is powerful, when it comes to naming a Chinese baby, I trust the ones that understand the culture"}. Moreover, some users reported that DouBao (a Chinese GenAI Tool) can generate aesthetically compelling visual imagery based on one's \textit{BaZi}, adding a unique layer of personalization which is shown in Fig.~\ref{fig: doubao}. These culturally embedded capabilities make Chinese GenAI particularly effective for supporting traditional divinatory frameworks.

\subsection{Social Interactions in Online Community}

\begin{figure}[h]
\centering
\includegraphics[width=1\linewidth]{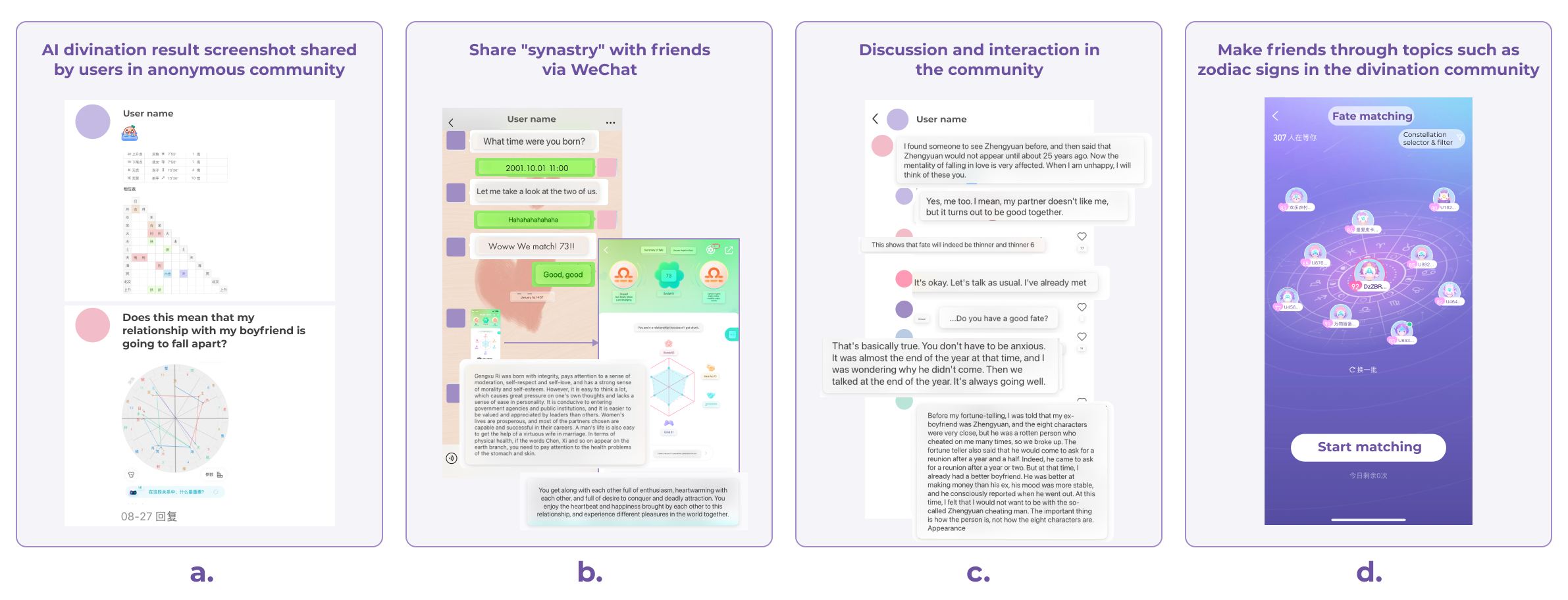}
\vspace{-0.6cm}
\caption{Screenshots of Sharing: (a) Anonymous sharing of AI-powered divination results (e.g., a natal chart interpretation from the Cece platform) on online communities like RedNote to initiate discussions. (b) Sharing Cece's interpretation of a compatibility chart ("synastry") with a friend via WeChat. (c) On user communities like RedNote, members discuss interpretations of specific emotional topics to find resonance and support. (d) Users connect and build friendships on Cece based on shared interests like astrology.}

\label{fig: comm}
\end{figure}

GenAI-based divination platforms have evolved into rich social environments where users not only consume personalized predictions but also engage in meaningful exchanges with others. Drawing on our netnographic observations and participant interviews, we identified that social sharing, belief reinforcement, and identity expression are central to how users interact with GenAI divination in online communities, as shown in Fig.~\ref{fig: comm}. We organize our findings into three interrelated themes: (1) content sharing and social motivations, (2) recipient dynamics and relationship-building, and (3) communal belief reinforcement and identity performance.

These interactions take place across a range of socio-technical platforms, each affording distinct modes of sharing. On private messaging platforms such as WeChat, divination results are exchanged in one-to-one or small-group conversations, particularly when users share synastry readings with partners or close friends. On interest-based social media platforms such as RedNote, users post AI-generated predictions under tags like “fortune” or “emotions,” which are then circulated through algorithmic recommendation to audiences with shared interests, creating semi-public discussions among strangers. Finally, dedicated divination applications such as Cece integrate social features directly into the platform, matching users based on “compatible” or “similar” astrological attributes and encouraging interaction through gamified mechanisms.

In these communities, users’ sharing of AI-based divination content transcends simple information exchange, becoming a process of transforming abstract AI outputs into social artifacts. This content—often presented as "visual charts" or "interpretative explanations"—is easily displayed and understood, providing a tangible basis for social interaction. While these posts are typically straightforward, they underpin a collaborative process of sense-making. For instance, one participant noted that sharing daily readings helped them "track their moods and outcomes over time" (P8), highlighting how sharing functions as a tool for self-reflection and personal narrative construction. Another user found that "seeing others' interpretations can give me new perspectives on my own results" (P15), which reveals the crucial role of collective interpretation in an individual's meaning-making. By inviting others to engage with these results, users turn a private AI divination experience into a public platform for social validation and analysis.

This process of turning private insights into a public platform for social analysis is defined by the specific audiences and relationships involved. The recipients of this shared information include romantic partners, individuals with similar interests, or friends. Users often discuss "synastry" (astrological compatibility) with these recipients, examining personal relationships and compatibility in romantic or friendship contexts. This creates an interactive and socially engaging environment where users exchange insights on relationships. As one participant shared, \textit{"I like to check compatibility between me and my friends and then share the results to get their reactions"} (P3). Another noted the appeal of sharing these insights within close social circles, commenting, \textit{"It's interesting to see how compatible I am with my partner based on our signs, and it opens up conversations about our relationship"} (P11).

The motivations for sharing divination results are diverse, but they primarily revolve around three key behaviors: seeking social validation, fostering collaborative interpretation, and building relationships. Many users share to gain social validation and affirmation from the community. As P16 shared, "I post to get reactions and to validate if others interpret the signs similarly," demonstrating a need for communal consensus on the AI's output. This act of sharing also serves as a way to "get other people's interpretations" (P13), providing an opportunity for users to collectively make sense of the results and find emotional support or alternative insights. For many users, the content itself is not as important as the opportunity to connect with others, highlighting a deeper intent of building friendships and relationships. Thus, the shared content functions as a social currency that facilitates connection, making the platform a space for meaningful social interaction.

\begin{figure}
    \centering
    \includegraphics[width=1\linewidth]{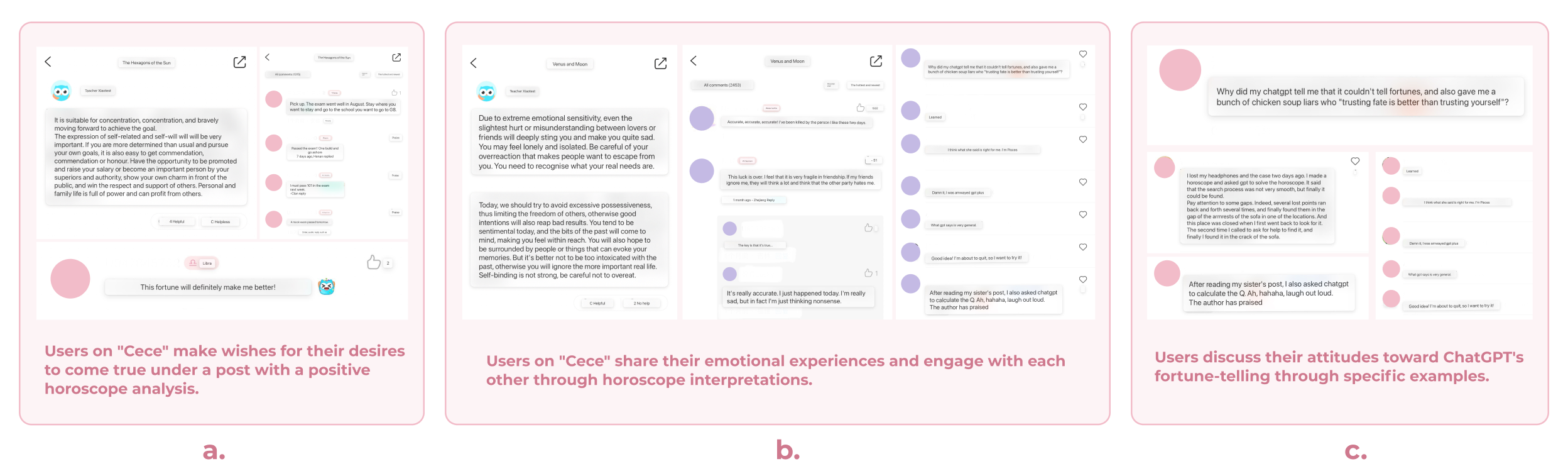}
    \vspace{-0.6cm}
    \caption{Screenshots in the community: (a) Users on Cece make wishes for their desires to come true under a post with a positive horoscope analysis. (b) Users on Cece share their emotional experiences and engage with each other through horoscope interpretations. (c) Users discuss their attitudes toward ChatGPT's fortune-telling through specific examples.}
    \label{fig: Screenshots}
\end{figure}

This collective behavior transforms GenAI-based divination communities into more than simple platforms for receiving predictions; they function as social spaces where personal beliefs are collectively validated and amplified. Within these communities, users frequently post and discuss their divination results, inviting others to participate in a form of distributed self-justification, where individual interpretations gain strength through communal endorsement, as some screenshots in Fig.~\ref{fig: Screenshots}. For example, in a thread discussing astrological predictions under the theme "Sun sextile Midheaven," one user enthusiastically shared their aspiration for academic success, stating, "I will pass the exam and advance!", a sentiment quickly echoed by other users responding with similar affirmations like "I'll pass the exam next week!" and "I'm advancing in my studies too!" These responses reinforce a collective belief in the accuracy and positive influence of divination, creating a feedback loop where users' hopes are bolstered by the optimism of others, ultimately strengthening individual convictions.

This phenomenon of collective belief reinforcement becomes especially poignant in emotionally charged contexts, such as discussions about relationships. In a post analyzing a challenging astrological alignment, "Venus square Moon," a user shared their distress, stating, "These past two days, the person I like has been driving me crazy." Others quickly responded with comments like, "This alignment makes me feel so fragile in friendships" and "It's incredibly accurate; something similar just happened to me today." These responses do more than validate individual experiences. They construct a shared narrative where users feel seen and understood, reducing feelings of isolation while reinforcing the perceived credibility of the GenAI-generated divination.

Furthermore, this process of belief reinforcement serves as a means of identity presentation within the community. By publicly aligning with popular interpretations, users not only solidify their personal beliefs but also craft an identity that resonates with the group's cultural framework. In one instance, a user humorously noted the limitations of GenAI, remarking, "Why did my ChatGPT say it can't do fortune-telling and gave me a bunch of 'believe in yourself' advice instead?" Other users joined in with lighthearted comments, echoing their own playful skepticism or curiosity about the AI's insights. Through such exchanges, users engage in identity performance that aligns with the community's shared norms, establishing a persona that combines both reliance on and critique of the AI’s capabilities.

The communal space created by these platforms allows users to participate in a unique form of "networked belief" construction, where individual convictions are collectively endorsed, nuanced, and sustained within a hybrid socio-technical environment. These GenAI-powered interactions extend beyond individual predictions, providing a collaborative arena for belief reinforcement, identity formation, and shared meaning-making. As users collectively negotiate and affirm their interpretations, the GenAI platform transforms into a communal space for social bonding, emotional support, and mutual validation, offering novel insights into how online communities mediate and reinforce belief systems.

\subsection{Common GenAI Fortune-telling Process}

\begin{figure}[h]
\centering
\includegraphics[width=1\linewidth]{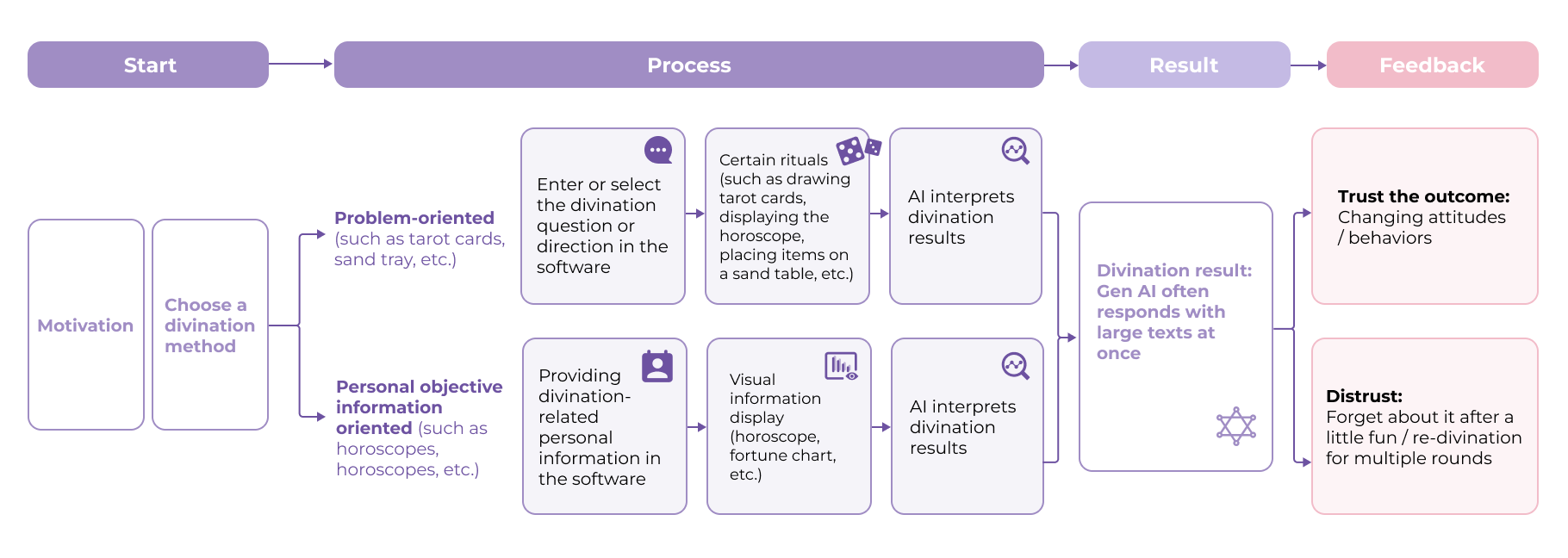}
\caption{Workflow of GenAI Divination: Starting from user motivation, selecting a divination method and process, and receiving AI-generated results, with user feedback varying based on trust in the response.}
\label{fig: workflow}
\end{figure}

Through interviews and ethnography, we explored the procedural stages involved in GenAI fortune-telling across various platforms. Participants were encouraged to provide in-depth descriptions and share screenshots of each step in the process. Based on data collected from 22 participants and online posts, we synthesized the workflow into four primary stages: Start, Process, Result, and Feedback, as shown in Figure \ref{fig: workflow}.

\subsubsection{Stage I: Motivation and Method Selection}

The process begins with users' diverse motivations and the selection of an appropriate divination method. Users come to GenAI divination with a range of intentions, such as seeking entertainment, finding psychological comfort, guiding minor decisions, or making informed choices about significant events. Depending on their specific goals, they choose a divination method that aligns with their needs. For instance, one interviewee, P2, was motivated by "anxiety and uncertainty in her environment" and often used AI late at night; therefore, she selected Cece because "she finds it accessible and available during late hours, which supports her need for immediate psychological comfort and short-term guidance." Another interviewee, P12, expressed a preference for drawing tarot cards herself, as she felt "the process retained the spirituality of divination." Consequently, she opted to use ChatGPT, as its general-purpose model allows her to upload images and questions freely, making the interaction feel more personalized and responsive.

\subsubsection{Stage II: Process}
\begin{figure}[h]
\centering
\includegraphics[width=1\linewidth]{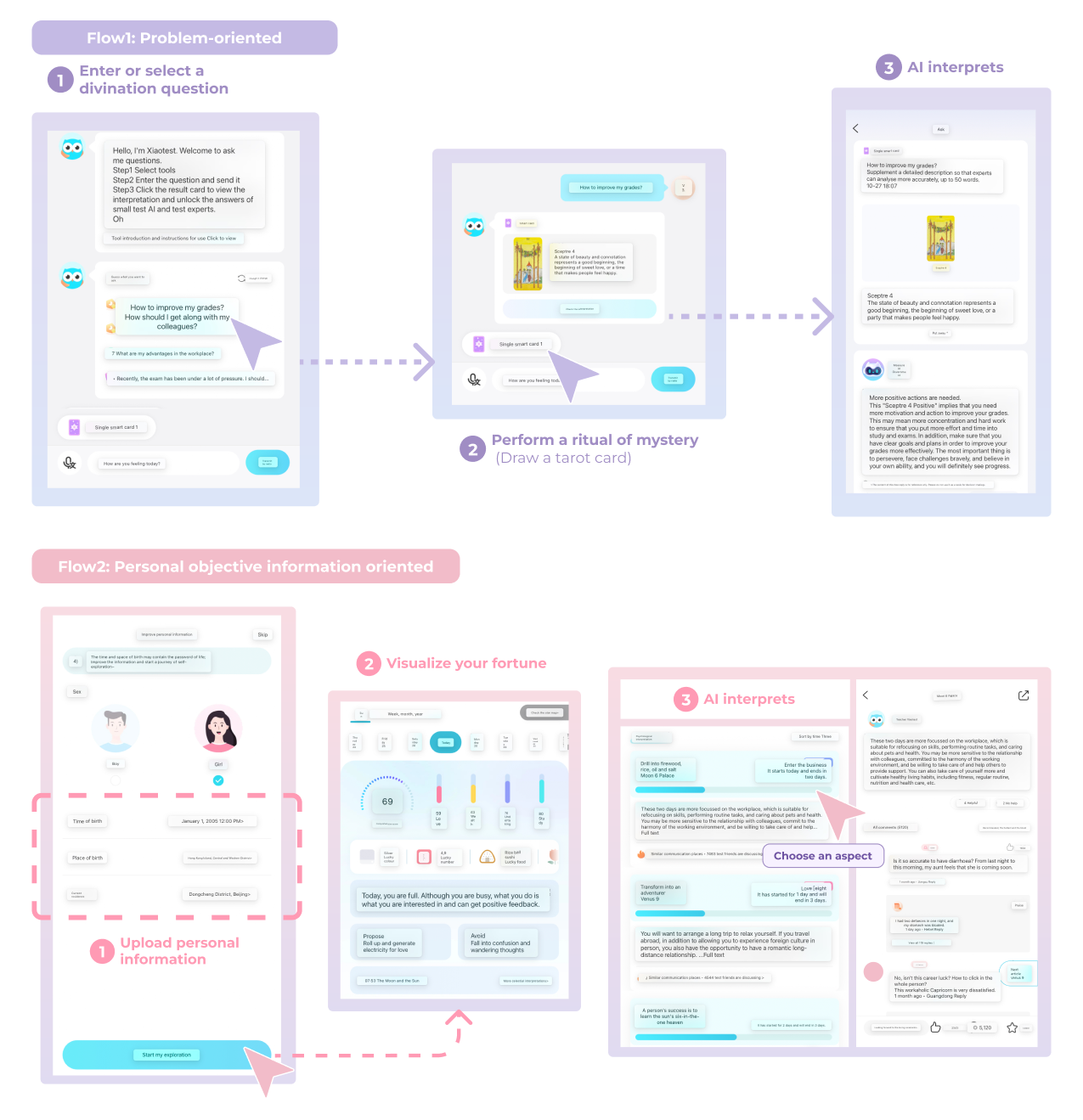}
\caption{Examples for problem-oriented and personal information-oriented flows. The Problem-Oriented Flow involves entering a question, performing a ritual, and receiving AI-generated results; the Personal Information-Oriented Flow requires users to input personal information, visualize their fortune data, and receive interpretive answers from GenAI.}
\label{fig: Process examples}
\end{figure}

After selecting a divination method, users proceed by focusing on their question or providing relevant personal information, leading to two primary types of process steps, which are illustrated in Fig.~\ref{fig: Process examples}.

\textit{Question-Oriented}. This approach is driven by a specific question from the user, like "Will I find a new job soon?" To make the experience more immersive, the system guides them through a "divination ritual." For instance, a tarot reading might involve virtually "shuffling" and "drawing" cards, or a sand tray divination might have them arranging digital objects. After the ritual, the GenAI interprets the results and provides an answer.

\textit{Personal Information-Oriented}. In contrast, this type of divination relies on objective personal data, unique to the individual, such as birth date and time for astrology, or facial photographs for physiognomy. Users need to provide these details to the system, and in some cases, certain information might already be pre-filled from their account registration. Unlike the question-oriented approach, this workflow usually lacks ritualistic elements; however, the application may enhance the experience with visual aids. For instance, if the user's constellation is Pisces, two cartoon fish icons might appear on the page. Following these introductory visuals, the GenAI then generates and displays the interpretation based on the provided data.

\subsubsection{Stage III: Prediction Outcome Perception}
This is the stage where GenAI responds with the divination outcome. Typically, the GenAI provides a detailed, text-heavy answer, sometimes spanning multiple paragraphs, to cover all aspects of the query. For instance, participant P2 appreciated the emotional comfort provided by GenAI's responses and shared, \textit{"GenAI tends to comfort me and rarely gives negative results. The overall score can affect my mood for the day, creating a sense of pre-judgment about life events."} However, some users find limitations in GenAI's interaction style. Participant P5 noted a lack of personalized engagement, saying, \textit{"The responses often feel templated, and when I ask follow-up questions, they can sometimes contradict the earlier answers."} These varied perspectives on GenAI's response quality and personalization will be discussed in more depth in the following sections.

\subsubsection{Stage IV: Feedback}
In the final stage, users assess the divination results and determine how to respond. For some, the reading resonates deeply with their current life circumstances, prompting them to adjust their mindset or behavior. For example, P6 noted that when GenAI predictions discouraged certain actions, she would \textit{"hesitate and think twice about its suggestion"} for that day, subtly aligning her behavior to the fortune's advice, especially for small daily choices. Conversely, others approach the results with skepticism, viewing the experience as lighthearted entertainment and moving on without further engagement. Notably, some users do not settle for a single divination session; instead, they return to the tool multiple times, subtly modifying their questions to see if the responses change or offer new insights, continuing until they feel satisfied with the answers. For example, participant P2 shared that, \textit{"If I feel there's a slight deviation, I'm not satisfied, I might recalculate."} Another participant, P6, echoed this practice, noting that she would sometimes ask the same question multiple times until the answers felt "right" or believable, reflecting a desire for responses that resonated with her current emotional state.

\section{Discussion}\label{sec:Discussion}
Our study situates GenAI divination within the broader conversations on human-AI interaction, cultural adaptation, and algorithmic trust. Whereas prior work on digital spirituality has primarily focused on Western contexts or generalized forms of algorithmic influence~\cite{TrustCrisisorAlgorithmBoost, DivinationanditsPotentialFutures}, our findings highlight how GenAI fortune-telling is embedded in culturally specific practices that reshape both individual and communal engagements with technology. In China, divination has long functioned as a way to manage uncertainty and sustain collective meaning. When mediated by GenAI, this practice becomes a site where technological systems are appropriated into existing cultural logics, producing new forms of psychological reliance and collective trust. Through RQs, we investigate both the ways users interact with GenAI divination tools and how they evaluate their effectiveness, revealing why AI-based predictions are selectively trusted or integrated into traditional practices~\cite{fu2023techno, TrustCrisisorAlgorithmBoost, DivinationanditsPotentialFutures}. 

\subsection{Key Findings on Trust, Agency, and Ritual in GenAI Divination}

The interaction between users and GenAI fortune-tellers represents a novel form of human-agent collaboration. Instead of functioning as a simple exchange with a passive tool, GenAI divination operates as a joint process of co-construction. In this practice, the user’s relational stance toward the AI becomes a defining factor. Whether they view the system as an objective data processor or a mystical intermediary directly shapes how spiritual meaning is generated. This collaborative partnership serves as the foundation for a broader, socially distributed process, where the insights co-created with the AI are sustained and contextualized through community discussion, repeated validation, and ritualized interaction.

From this perspective, the integration of GenAI into divination marks not just a technological upgrade but a paradigmatic shift in how spiritual knowledge is accessed, interpreted, and trusted. Rather than simply digitizing existing rituals, GenAI reframes the divinatory encounter into a more transactional, hyper-personalized, and algorithmically mediated experience. This transformation generates new possibilities, expanded accessibility, customizable content, and perceived objectivity, but also raises pressing concerns about emotional depth, ethical responsibility, and the commodification of uncertainty~\cite{Fisher2023CenteringTH, Feng2024FromHT}.

A widely discussed psychological theory known as "automation bias" suggests that people tend to trust computer-generated outputs, even when they may be flawed or biased~\cite{binns2018s}. This may help explain why many participants perceived GenAI’s algorithmic neutrality as more trustworthy than human fortune-tellers, who may, consciously or unconsciously, tailor their responses to meet social expectations. Participant P8 observed, “AI feels more genuine because it isn’t influenced by what I want to hear.” This perception aligns with broader sociotechnical discourses around algorithmic authority and “machine objectivity”~\cite{TrustCrisisorAlgorithmBoost}, where users equate lack of emotion with epistemic purity~\cite{lustig2016algorithmic}. Yet, this trust in AI’s neutrality may obscure the fact that GenAI responses are ultimately shaped by training data, prompt templates, and system biases, rather than spiritual insight. What is framed as “objective” may in fact be unacknowledged algorithmic authorship~\cite{zhou_tell_2026}, detached from the cultural intuition and symbolic literacy that traditional practitioners cultivate over years of practice~\cite{shin2021effects}.

Another significant shift lies in the user-driven customization GenAI enables. Rather than submitting to a practitioner’s guidance or symbolic framing, users can dictate topics, pace, and interpretation scope, an inversion of the traditional fortune-telling relationship. Participant P7 reflected, “With AI, I can generate results on any topic I want without worrying about what a fortune-teller might think.” Moreover, the way users exercise this agency is deeply influenced by their individual psychological traits and emotional states. For participants with higher levels of introversion or social anxiety, the GenAI interface creates a psychologically safe zone that traditional face-to-face divination often lacks. By removing the pressure of social performance and the fear of judgment from a human practitioner, the AI allows these users to engage in deeper self-disclosure. In this private setting, the technology functions less as a mysterious oracle and more as a reflective mirror, enabling users to explore their vulnerabilities without the burden of interpersonal conversations.

While this user agency appears empowering, it also risks flattening the divinatory experience into a preference-based consumption model, where users selectively engage with content that aligns with existing beliefs, even generating for many times to get "a good fortune", thereby reinforcing confirmation bias. Moreover, as we discuss in Section~\ref{sec:4.1.1} this agency allows users to bypass the ethical boundaries inherent in traditional practices; unlike human practitioners who may refuse to answer harmful or taboo queries, GenAI enables users to pursue sensitive topics, such as life and death. Another risk is the emotional shallowness of GenAI interactions. Traditional fortune-telling often incorporates ritual, empathy, and atmosphere, elements that lend depth to the experience and create a therapeutic, immersive environment. Many users report that GenAI, while efficient, lacks this emotional resonance. Participant P6 described it as feeling \textit{"too logical or formulaic, missing the mysterious aura and spiritual ambiance of traditional divination."} This lack of depth can lead to a transactional, impersonal experience that may fall short of fulfilling the complex emotional needs that users bring to divination.

Consequently, the internal psychological safety and unsatisfied emotional needs also shape how users extend their private reflections into social interactions~\cite{zeng_ronaldos_2025}. We found that a desire for social connection often drives the decision to share AI-generated results. For these users, the algorithmic prediction transforms into a "social object"~\cite{simon2010participatory} that facilitates interaction. By sharing a "neutral" AI fortune, they can initiate conversations, seek empathy, or validate their feelings within their community while maintaining a comfortable emotional distance. Thus, the GenAI system not only acts as a private sanctuary for the introverted but also a bridge for those seeking to reconnect with their social circles.

To improve GenAI fortune-telling, future systems should support emotionally adaptive, dialogue-based interactions rather than static templates~\cite{jiang_hear_2026}, incorporate immersive elements that restore the ritualistic and affective depth of traditional practices, and embed ethical safeguards to prevent harm in sensitive scenarios. Together, these strategies call for a reimagining of GenAI divination not as a tool of replacement, but as a new cultural interface, one that respects the spiritual textures of traditional practices while innovating with care and accountability.

\subsection{Key Findings for Collective Belief and Identity Formation in GenAI Divination Online Communities}

Our ethnography extends this lens to show that GenAI fortune-telling not only mediates personal meaning-making but also fosters new forms of digital community and social connection among those sharing similar beliefs~\cite{Astudyof, Applicationof}. This process could be driven by classic sociological concepts like "Social Identity Theory"~\cite{tajfel2001integrative} and "Collective Effervescence"~\cite{durkheim2016elementary}. "Social Identity Theory" suggests that individuals define themselves by belonging to social groups, drawing self-esteem from shared beliefs. 

In traditional divination, the human fortune-teller acts as the sole authority and emotional anchor, providing both prediction and empathetic interpretation. GenAI, however, provides the "raw data" of prediction but often lacks the "human touch." Our findings suggest that online communities emerge to fill this gap; the authority previously held by the single human diviner is now displaced onto the community. GenAI platforms like Cece serve as social spaces where individual beliefs are collectively validated. When a user posts an AI-generated astrological reading, the result acts as a shared artifact that bridges the gap between strangers, allowing them to coordinate their meanings and emotions. Unlike the vertical authority of a traditional diviner, the authority in GenAI communities is distributed and horizontal. Users do not rely solely on the AI (the digital diviner); instead, they rely on the community to interpret, debug, and contextualize the AI's output.

This collaborative validation forms a powerful belief loop. When a user posts a GenAI prediction about passing an exam based on a positive reading, other members quickly echo this belief. This synchronized interaction is more than information exchange; it becomes a form of digital ritual. Through this ritual, members experience "Collective Effervescence"~\cite{durkheim2016elementary}, a shared energy that strengthens group cohesion. This shared positivity strengthens group cohesion and reinforces each member's trust in the divination results.

This belief reinforcement process also serves as a means of "identity presentation"~\cite{goffman2023presentation}, where users align their interpretations with community-supported norms, building a sense of belonging. By sharing interpretations and aligning with common views, users craft a public persona in Goffman's Dramaturgy that resonates with the group’s shared values, combining both reliance on and a playful critique of GenAI’s capabilities~\cite{Voorneveld2024ExploringIP}. The deeper implication is that GenAI platforms act as both belief amplifiers and community incubators~\cite{festinger1954theory}. The design of these platforms, including features for sharing and commenting, is not value-neutral. They create a powerful social validation system that can efficiently turn individual, uncertain beliefs into collective, endorsed facts. 

Ultimately, our results reveal that these platforms function as spaces for collaborative hermeneutics where interpretation becomes a group negotiation. Discussions often involve empathetic repair, where members contextualize standardized AI responses, such as reframing a generic prediction of "misfortune" into constructive life advice, to fill the emotional void left by the software. Simultaneously, users engage in technical critique, such as debating the optimal prompts to correct the AI's miscalculation of Lunar dates or BaZi charts. Through these interactions, the community actively reshapes algorithmic authority, transforming opaque technical outputs into culturally intelligible and collectively owned narratives.

\subsection{Design Implications}

\subsubsection{Adapting GenAI to Cultural Beliefs}
Our research provides a contemporary example of "techno-cultural domestication"~\cite{silverstone200512}. This is especially evident in the cultural embedding of algorithmic systems. This theory suggests technology is not passively accepted. Instead, users actively integrate and adapt it to fit local cultural values and practices. We found that users strongly prefer AI models capable of performing localized forms of divination. For instance, they use tools like DeepSeek, which are trained on Chinese corpora, for complex practices like BaZi or Ziwei Doushu. Participants noted that non-localized models struggle with the intricate cultural symbols found in Chinese metaphysics. In contrast, DeepSeek produces more culturally authentic and contextually appropriate interpretations.

Thus, the future LLM model for fortune-telling could benefit from domain-specific fine-tuning. As the method in CulturePark~\cite{li2024culturepark}, developers should integrate specific metaphysical corpora (e.g., ancient texts like the \textit{I Ching} or \textit{BaZi} calculation rules) into the model's training data or use Retrieval-Augmented Generation (RAG)~\cite{lewis2020retrieval} to ground answers in authentic cultural logic. Furthermore, we suggest the design of "Cultural Personas" as mentioned in Pataranutaporn's work~\cite{pataranutaporn2023influencing} to replace the default "neutral assistant" persona commonly. Unlike productivity tools, where a neutral tone implies objectivity, in divination, neutrality can be perceived as "soulless." A persona that mimics a traditional elder, using semi-classical Chinese semantics and a tone of empathetic authority, better matches the user's mental model of a practitioner. This aligns with recent work in educational agents, which shows that "cultural congruency" significantly enhances user trust and information uptake.

\subsubsection{Balancing Objectivity and Mysticism}
Our study also reveals the prevalence of "cognitive polyphasia"~\cite{moscovici2008psychoanalysis} among users. This theory describes how individuals can simultaneously hold and apply contradictory belief systems, such as scientific rationality and metaphysical beliefs. This concept explains the core epistemic tension we observed. On one hand, users perceive AI as an objective and neutral product of science. They believe it is not influenced by a desire to please them. On the other hand, they eagerly seek metaphysical comfort and guidance from this same scientific tool. This tension is clear in their behavior. They appreciate the AI's objectivity yet repeatedly push it to produce results that meet their cultural and emotional expectations. This dynamic upends the simple model of the user as a purely rational agent.

Addressing this tension requires interaction designs that introduce "meaningful inefficiency." We propose incorporating "Digital Rituals" via artificial latency. For example, Buçinca~\cite{buccinca2021trust} introduced "cognitive forcing functions" (deliberate friction) to force users to think analytically and reduce over-reliance on AI. However, our implication flips this logic: we suggest using similar friction mechanisms (e.g., visualizing a 10-second star blink animation) not to trigger analytical thinking, but to induce psychological solemnity. Unlike productivity AI, where speed is paramount, here the waiting time serves as a liminal space for users to suspend disbelief. Additionally, to create the "safe space" for meaning-making mentioned in Section 5.1.2, systems could offer explicit Mode Switching. Users could choose between a "Rational Analysis Mode" (data-driven) and a "Spiritual Guidance Mode" (metaphor-driven). This gives users agency over the framing of the advice, thereby reducing the cognitive dissonance between their scientific education and metaphysical needs.

\subsubsection{GenAI as a Tool for Emotional Support}
Finally, our research highlights the key role of AI in users' emotional and psychological regulation. This phenomenon can be explained by the "Barnum Effect"~\cite{forer1949fallacy} and "confirmation bias"~\cite{wason1960failure}. GenAI's outputs are often vague and general. This quality allows them to provide effective emotional comfort as users perceive them as personally accurate. Stronger evidence comes from a unique interaction pattern we observed. Users who are unhappy with a result will repeatedly rephrase their questions until they receive a desirable answer. The user's goal is not to find an objective truth. It is to construct a narrative that provides emotional relief in times of uncertainty. This reframes how we should define success for such AI systems. The system's value is not measured by its predictive accuracy but by its effectiveness in meeting users' psychological needs.

Similar to the narrative goals in \textit{Worlding with Tarot}~\cite{Michelson2024WorldingWT}, GenAI divination can be designed as a tool for meaning-making and emotion support rather than just fortune-telling. Co-constructive interfaces will be a better approach. Instead of allowing users to simply regenerate results until they get a "good fortune" (a passive loop), systems could implement "Interactive Reframing." For example, if a user receives a negative prediction, the chatbot could prompt: \textit{"How does this result align with your current worries? What action could change this outcome?"} This shifts the interaction from passive fatalism to active emotional regulation.

Moreover, identifying the user's specific divination goal is another critical strategy. We recommend implementing an "Intent-Aware Calibration" module or Linguistic Marker Analysis. This allows the system to distinguish between users seeking playful entertainment (who prefer witty, mystical language) versus those seeking decision support or deep emotional validation (who need clear, empathetic grounding). For instance, if the system detects high-anxiety markers (e.g., repeated use of "scared," "fail"), it should automatically shift its persona from a cryptic oracle to a supportive guide. This mimics the cold reading skill of a human practitioner, assessing the client's mental state before delivering a verdict, thereby ensuring the AI's tonal strategy aligns with the user's immediate psychological needs.  Context-awareness is also important for emotional support. When queries involve high-stakes topics (e.g., health crises, suicidal ideation), the system should detect these intent signals and provide ethical hints.

\subsection{Limitations}
\subsubsection{Demographic Limitations}
The study's participant demographics were predominantly limited to young, educated Chinese individuals, with a gender imbalance (14 female, 8 male) and concentration in design-related majors (5 out of 22 participants). These limitations arose primarily from our recruitment methods through university networks and social media platforms popular among young adults in China. While this approach enabled the efficient recruitment of participants actively engaged with GenAI fortune-telling tools, it inadvertently excluded significant demographic segments, for instance, participants with non-academic backgrounds and seniors. The absence of these perspectives means our findings may not fully represent the broader spectrum of GenAI fortune-telling users, particularly those who might approach these tools with different cultural, educational, or socioeconomic contexts. For instance, while some Chinese users describe GenAI divination as offering a heightened sense of emotional safety than a fortune-teller, users in other regions may feel insecure about issues such as data privacy or algorithmic transparency. Future research would benefit from actively recruiting participants from these underrepresented groups to provide a more comprehensive understanding of GenAI fortune-telling adoption and use across diverse communities.

\subsubsection{Experience Level Distribution}
The study's participant experience levels showed an uneven distribution, with 6 frequent users compared to 13 occasional users and 3 one-time users. This skewed distribution emerged largely from the nascent nature of GenAI fortune-telling technology and its recent introduction to the market, making it challenging to find long-term, experienced users. Our research focused on this particular distribution as it reflected the current adoption pattern of GenAI fortune-telling tools, where most users are still in the exploratory phase of engagement. However, this approach meant we could not thoroughly examine the long-term effects, evolved usage patterns, or deeper insights that might come from sustained engagement with these tools. The limited number of frequent users in our sample restricted our ability to understand how prolonged use might influence trust development, decision-making patterns, or the integration of GenAI fortune-telling into daily routines. We were unable to include more frequent users partly due to the timing of our study coinciding with the early stages of GenAI fortune-telling adoption, and partly due to the difficulty in identifying and recruiting users who consistently engage with these tools over extended periods. This gap in our research suggests that longitudinal studies tracking users' evolving relationships with GenAI fortune-telling tools would be valuable for future research.

\subsubsection{Platform Coverage}
The study's platform coverage primarily focused on ChatGPT, DeepSeek, and Cece, with a few participants having experience with alternative GenAI fortune-telling platforms. This concentrated coverage reflected the market dominance of these platforms in China during our research period. While this focus allowed us to conduct an in-depth comparison between a general-purpose GenAI tool (ChatGPT) and a specialized divination application (Cece), providing rich insights into how users adopt different types of GenAI platforms for fortune-telling purposes, it also limited our understanding of the broader GenAI fortune-telling landscape. Although the study touches on emerging tools like Ziwei Doushu AI, Quin AI, and TianGong AI, these represent only a small part of the broader landscape of GenAI divination tools. This limitation arose from the rapid evolution of the GenAI field, making it challenging to include all relevant platforms during our research period~\cite{Voorneveld2024ExploringIP}. Future research would benefit from examining a wider range of GenAI fortune-telling platforms to understand how different technical approaches and design philosophies influence user experiences.

\subsubsection{Self-Report Long-term Data}
While our study relied on self-reported data through interviews, we did incorporate some direct observations by asking participants to demonstrate their interactions with GenAI fortune-telling tools during the interview. However, such longitudinal data was not retained by all participants and was generally unavailable for our analysis due to personal privacy habits or platform data clearance. Due to privacy concerns, participants might have withheld sensitive information about personal questions asked to GenAI fortune-telling tools and be unwilling to share it with us. Consequently, a key limitation is the absence of these most genuine forms of older interaction data. Participants' accounts may have been influenced by social desirability bias, particularly regarding their reliance on GenAI for decision-making and spiritual practices. Memory-related biases could have affected their recall of historical experiences, especially for occasional users. Although digital ethnography helped validate current usage patterns, these online scenarios might not fully capture natural, unprompted interactions with these tools. While the combination of interviews and ethnography provided valuable insights, the study could have benefited from systematic observational data over time oraccess to longitudinal or historical interaction logs to further validate self-reported experiences. Future research should avoid the "observer effect" and explore implementing historical content studies where participants document their real-time interactions with GenAI fortune-telling tools, or develop privacy-preserving methods to analyze user-AI conversation logs, providing more authentic insights into user behaviors and decision-making patterns.

\section{Conclusion}\label{sec:Conclusion}
Through 22 semi-structured interviews with GenAI fortune-telling users and a three-week digital ethnography with 1,842 posts, we explored how individuals engage with AI-based divination tools, how they perceive and trust the generated results, and how these tools influence their beliefs and interactions within communities. Participants selected GenAI methods that aligned with their personal motivations, using them for entertainment or psychological comfort, or as supplementary guidance for major life decisions. While some valued GenAI’s objective responses, others criticized its lack of mystical elements and ethical discretion, particularly when addressing sensitive or potentially harmful questions. Our findings suggest that GenAI fortune-telling tools should be enhanced to offer more personalized and interactive experiences, incorporate immersive design elements to create a ritualistic atmosphere, and implement ethical safeguards to responsibly manage high-stakes inquiries, thus fostering both engagement and user trust.

\section{Acknowledgments}
\add{We acknowledge using ChatGPT to assist with proofreading to improve the readability and language of this manuscript. The authors reviewed all AI-generated suggestions, made manual revisions, and retain full responsibility for the final content.}

\bibliographystyle{ACM-Reference-Format}
\bibliography{references}

\appendix

\section{Interview Questions}
\label{app: questions}

\subsection{Background Information}

\noindent How long have you been involved in fortune-telling? What methods do you usually use?

\noindent How long have you been using AI for fortune-telling? Which AI platform(s) do you use?

\noindent Do you use paid AI fortune-telling platforms?

\noindent How much do you pay or are willing to pay per month?

\noindent How did you initially come across generative AI in the context of fortune-telling?

\noindent Why did you choose to use AI for fortune-telling?

\noindent Were there any surprises when you first started using generative AI for fortune-telling?

\noindent What is your educational background? What was your major? What is your current occupation?

\subsection{Workflow and User Experience}

\noindent Can you describe your typical process for an AI-driven fortune-telling session? 

\noindent What input do you provide to the AI, and how is the output presented?

\noindent Compared to traditional fortune-telling, what differences do you observe in your AI fortune-telling process?

\noindent Which method do you prefer? Why? 

\noindent What aspects of AI fortune-telling attract you?

\noindent Are there any parts of the process that you find particularly challenging?

\noindent In your opinion, what improvements could generative AI make to enhance the fortune-telling experience?

\noindent What meaning or impact do AI-generated fortune-telling results have for you?

\subsection{Challenges and Misconceptions}

\noindent What common misconceptions or biases about AI fortune-telling do you think exist in society?

\noindent Have these misconceptions affected your experience?

\noindent How do you feel about the accuracy of AI-generated fortune-telling results? Why?

\noindent Compared to traditional fortune-telling, which do you find more accurate? Why?

\subsection{Ethical Issues in AI Fortune-Telling}

\noindent What are your thoughts or concerns about the ethics of AI fortune-telling?

\noindent Do you think generative AI can replace traditional human fortune-tellers? If not, what are the gaps?

\noindent Do you have privacy concerns when using AI for fortune-telling? 

\noindent How important are data security and privacy protection to you?

\noindent Would you have any reservations if AI fortune-telling involved more personalized information?

\subsection{Cognitive and Psychological Impact}

\noindent How does generative AI fortune-telling affect your emotions or mindset? 

\noindent Does it make you more inclined to believe in the results, or do you become more skeptical?

\noindent Do you think AI fortune-telling can understand your personal emotions and needs? 

\noindent How do you feel about this kind of “understanding”?

\noindent Do AI-generated fortune-telling results influence your life, behavior, or decision-making? Could you provide an example?

\subsection{Social Interaction}

\noindent During interactions with generative AI, do you feel it acts as a fortune-teller or more like a tool?

\noindent Have you ever used AI fortune-telling with friends or in a group setting? If so, could you provide an example?

\subsection{Future Prospects and Potential}

\noindent What is your view on the development potential of AI fortune-telling? 

\noindent What new features or experiences would you like it to offer?

\noindent In what ways do you think AI fortune-telling could be applied in psychological support or other fields?

\subsection{Additional Questions}

\noindent Do you think generative AI fortune-telling aligns with your belief system or cultural background? Why?

\noindent Has generative AI fortune-telling content on social media influenced your views on your own life or the lives of others?

\end{document}